\documentclass[%
 reprint,
nofootinbib,
nobibnotes,
 amsmath,amssymb,
 aps,
floatfix,
]{revtex4-2}

\usepackage{amsmath}
\usepackage{physics}
\usepackage{graphicx}
\usepackage{dcolumn}
\usepackage{bm}
\usepackage{subfigure}
\usepackage{xcolor}

\begin{document}


\title{Towards realization of universal quantum teleportation using weak measurements}
\author{Vivek Balasaheb Sabale$^{1}$}
\email{sabale.1@iitj.ac.in}
\author{Atul Kumar$^{1}$}
\email{atulk@iitj.ac.in}
\author{Subhashish Banerjee$^{2}$}
\email{subhashish@iitj.ac.in}
\affiliation{$^{1}$ Department of Chemistry \\ Indian Institute of Technology Jodhpur, 342030, India}
\affiliation{$^{2}$ Department of Physics \\ Indian Institute of Technology Jodhpur, 342030, India}

\begin{abstract}
In this manuscript, we analyze universal quantum teleportation in the presence of memory or memory-less dynamics with applications of partial collapse measurement operators. Our results show that the combined effects of memory or non-Markovianity and weak measurements can lead to universal quantum teleportation (UQT). Our study involves noise models of physical importance having characteristic Markovian and non-Markovian regions allowing one to observe a transition in quantum properties as one switches from non-Markovian to Markovian dynamics. For this, we characterize the effects of different types of non-Markovianity for efficient  UQT both due to retention of correlations for a longer duration and due to information backflow. We further analyze memory effects arising from a correlated channel with or without weak measurements. Interestingly, our analysis for a correlated amplitude damping channel shows that memory effects are of significant advantage to minimize the fidelity deviation. The presence of weak measurements further enhances the realization of UQT in the presence of memory. The ability of memory effects in achieving zero fidelity deviation at non-zero time is interesting and of experimental importance.
\end{abstract}
\maketitle

\section{\label{sec:level1}Introduction}
The successful experimental demonstration of the violation of Bell's inequality \cite{PhysRevLett.49.91,PhysRevLett.47.460,aspect1999bell,PhysicsPhysiqueFizika.1.195} proved the superiority of quantum ideas and laws to explain nature's actions, ending the debate over whether quantum mechanics or classical mechanics is at the core of nature's deeds. This experimental demonstration further encouraged researchers to explore the field of quantum information to harness nature's potential for handling information and develop superior technologies. The necessity of stable quantum systems is an undeniable need for realizing such quantum technologies. Generally, all quantum systems are susceptible to the external environment \cite{banerjee2018open,breuer2002theory,lidar2019lecture}, which is the primary obstacle to implementing these quantum technologies. There have been a number of impressive technological advancements in this field, a prominent example of which is the experimental confirmation of quantum teleportation (QT) \cite{bouwmeester1997experimental,PhysRevLett.80.1121,pirandola2015advances}. The potential use of QT \cite{gisin2007quantum,pirandola2015advances,fuchs2000quantum,peres2004actually,vaidman2006another,zeilinger2018quantum} in the technological advancement of communication was a major propelling force in making it a reality. The experimental demonstrations, on a single arbitrary qubit state, in \cite{PhysRevLett.80.1121,bouwmeester1997experimental} demonstrated the viability of the concept proposed in  \cite{PhysRevLett.70.1895} by destroying and reconstructing the quantum state using classical information and the non-local correlation of the EPR \cite{einstein1935can} (Einstein Podolsky Rosen) channel.\par

In QT, the input state is teleported to an arbitrarily distant location without time delay. However, it is not superluminal communication, as one needs classical information from the sender's end to make the QT successful with probability 1. To determine teleportation success, the overlap between the input state from the sender (Alice) and the output state received by the receiver (Bob) is calculated and called fidelity \cite{PhysRevA.62.024301}. The fidelity depends on nonlocal correlations of the EPR channel; a decrease in such correlations leads to a decrease in the perfect overlap of input and output states in the QT protocol \cite{wang2020quantum,10.5555/2230916.2230917,10.5555/3179473.3179475}. In addition, fidelity may show fluctuation due to a number of reasons ranging from inherent properties of resource state to environmental noise or input states used for QT. A QT protocol free from such fluctuations is called Universal Quantum Teleportation (UQT) \cite{PhysRevA.86.062317,bang2018fidelity}. The increased fluctuations in fidelity imply a departure from UQT- fidelity deviation \cite{PhysRevA.86.062317,bang2018fidelity} is used as a measure to quantify such departures. Therefore, in addition to fidelity, we emphasize on fidelity deviation to understand the impact of Markovian and non-Markovian noise of physical importance on QT and UQT. In the case of a noiseless EPR channel, a perfect input and output state overlap is possible, and fidelity deviation will be zero. However, in a real scenario, all quantum systems are open \cite{breuer2002theory,banerjee2018open,lidar2019lecture} and constantly interact with their immediate surroundings, which causes deteriorating effects on entanglement and non-local correlations \cite{PhysRevA.99.042128}. In that case, we always get a fidelity lower than the optimum and fidelity deviation greater than zero \cite{PhysRevA.66.022316}. \par

\begin{widetext}
\begin{figure*}
\includegraphics[width=0.9\textwidth]{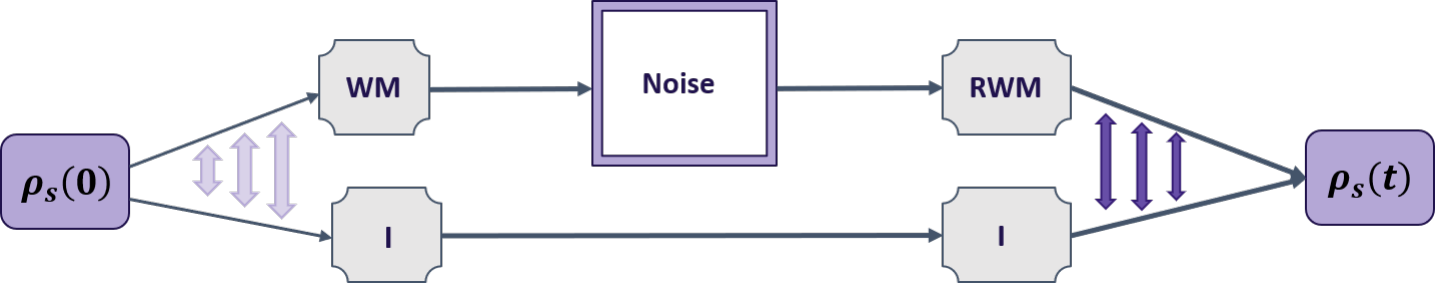}
\caption{WM and RWM protocol implemented in case of ADC, PLN, OUN, RTN noise. WM and RWM imply the weak measurement and its reversal operation performed on the first qubit, and I stands for identity operation. The verticle arrows represent the action of weak measurement in enhancing correlations.} 
\label{fig:wm_protocol}
\end{figure*}
\end{widetext}

This decay of correlation can be addressed using entanglement distillation, decoherence free subspace \cite{PhysRevLett.81.2594,PhysRevLett.84.4733}, quantum error correction codes \cite{PhysRevA.52.R2493,PhysRevA.54.1098,PhysRevA.55.900} and weak measurements (WM) \cite{PhysRevLett.60.1351}. In this article, we revisit the applications of weak measurements to analyze effects of non-Markovian noise and memory on QT and UQT. WM being a partial collapse measurement \cite{lahiri2021exploring} allows to recover the initial state by performing a weak measurement reversal operation \cite{PhysRevLett.97.166805}. The reversal operation provides probabilistic success in reconstructing the initial state and therefore results in restoring the nonlocal correlations. The potential use of WM \cite{PhysRevLett.60.1351} and reversal of weak measurement (RWM) \cite{Kim:09} in presence of Markovian, non-Markovian and correlated channels to achieve UQT is presented in this work. For this, we perform a detailed study on UQT with different noise models of physical importance such as Power Law Noise (PLN), Ornstein-Uhlenbeck noise (OUN), Random Telegraph Noise (RTN), and a correlated amplitude damping noise. The interacting environment of a quantum system invokes Markovian (memoryless) or non-Markovian (memory) dynamics \cite{bylicka2014non,RevModPhys.88.021002,li2019non,milz2021quantum} based on the presence or absence of a clean separation of system-environment time scales. One of the well known identifiers of non-Markovian behavior are the information back-flow from the environment to the system \cite{PhysRevLett.103.210401,chruscinski2011measures} or departure from CP-divisibility criteria \cite{PhysRevLett.105.050403}. To identify the information back-flow, one relies on measures based on trace distance and its time derivative. The non-Markovianity of non-Markovian AD noise, non-unital in nature,  and Random Telegraph Noise (RTN), which is unital, \cite{kumar2018non,lalita2023harnessing} are captured by it. There is also a weaker condition of non-Markovianian dynamics, which arises from the retention of correlation even in CP-divisible processes; for example, the unital noise channels, the modified Ornstein-Uhlenbeck noise (OUN) and Power Law noise (PLN) \cite{utagi2021non,shrikant2018non}. We present the advantages of both types of non-Markovianity in QT protocol. FIG.  \ref{fig:wm_protocol} represents the schematic model to analyse effects of these channels on nonlocal correlations and their dynamics its relation to UQT. \par
The paper aims to present a study of time-dependent dynamics of quantum correlations and effects of different types of non-Markovianity in satisfying the UQT conditions. Further, it also presents the ability of WM and RWM to increase the effectiveness of even a less entangled state as a resource for UQT. In fact, our results show that an entangled state with concurrence as less as $0.19$ can be used as a resource to achieve the same efficiency in UQT as a maximally entangled Bell state in presence of noises studied in this article. We further study the effect of memory to show that it enhances the efficiency of QT protocol, making it one step closer to UQT. Our analysis for correlated amplitude damping channel (CADC) \cite{PhysRevA.67.064301} possessing memory shows that memory effects and weak measurements can lead to control over fidelity deviation in achieving a value of zero at different times. Also, in the case of non-Markovian noise, the WM and RWM can cause enhancement in indicators of non-Markovianity, such as trace distance and its time derivative \cite{PhysRevA.81.062115}.\par

The paper is organized into sections, starting with basic ideas and concepts in section \ref{sec:level1i}. It contains a discussion about an entanglement measure, i.e., concurrence, fidelity, fidelity deviations \cite{PhysRevA.86.062317,Ghosal_2020}, and quantum channels. We explore the ideas of WM, non-Markovianity, correlated channels and their effects on concurrence \cite{PhysRevLett.78.5022}, coherence \cite{PhysRevLett.113.140401}, QT and UQT. In section \ref{sec:level4}, we also analyze the influence of non-Markovian channels using indicators like trace distance and its time derivative. We further present the memory effects introduced via correlated amplitude damping and its impact on fidelity and fidelity deviations in section \ref{sec:levelWm}. We further summarize the results obtained for multiple quantum channels in TABLE \ref{tab:noise}.

\section{\label{sec:level1i} Basic ideas and concepts}
Entanglement \cite{RevModPhys.81.865,popescu2006entanglement,peres1998quantum,friis2019entanglement} is a fundamental property that distinguishes and elevates quantum mechanics above classical mechanics in its applicability to understanding nature. The fundamentals of quantum mechanics permit composite quantum systems to exist in states that cannot be decomposed tensorially into the states of their constituent quantum subsystems. This allows quantum systems to manifest properties like non-locality \cite{bohm1977intuitive,popescu2014nonlocality,RevModPhys.86.419,PhysRevA.54.2685} and entanglement. Experimentally, it has also been observed that quantum particles can exist in an entangled state, and an entangled state could be used as a resource in quantum information and computation \cite{nielsen2001quantum,wootters1998quantum}. For example, entangled resources have been efficiently used as resources in several key protocols including QT, dense coding \cite{PhysRevLett.69.2881}, quantum cryptography \cite{RevModPhys.74.145}, and many others \cite{ursin2007entanglement,yin2017satellite,pauls2013quantum}. The transition in entanglement study as a resource has also increased the need to quantify resource capability present in quantum states. Using a measure of quantum entanglement, it is possible to comprehend the quantum state's performance in above mentioned protocols.  For our purpose, we limit our discussion to concurrence \cite{PhysRevLett.80.2245} which is a measure of entanglement in two qubit systems. In addition to concurrence, we also analyze a measure of coherence known as $\textit{l}_{1}$-norm \cite{RevModPhys.89.041003,PhysRevLett.113.140401}. In order to facilitate the analysis of results obtained, we first briefly describe fundamental concepts used to characterize and analyse different Markovian and non-Markovian noisy channels for studying information back-flow, fidelity and fidelity deviation. 
\subsection{Concurrence}
One of the most fundamental and promising quantifiers of degree of entanglement in two qubit systems is concurrence \cite{PhysRevLett.80.2245} where concurrence of a two-qubit quantum state $\rho$ is defined as:
\begin{equation}
    C(\rho)=max\{ 0, \lambda_{1}-\lambda_{2}-\lambda_{3}-\lambda_{4}\},
\end{equation}
Here $\lambda_{1},\lambda_{2},\lambda_{3},\lambda_{4}$ are square roots of eigenvalues of the matrix $\rho \Tilde{\rho}$, and $\Tilde{\rho}=(\sigma_{y}\otimes\sigma_{y})\rho^{*}(\sigma_{y}\otimes\sigma_{y})$ is a state flipped via action of Pauli-Y matrix and $\rho^{*}$ is complex conjugate of a state $\rho$. 

\subsection{Coherence Measure}
Quantum coherence \cite{RevModPhys.89.041003,PhysRevLett.113.140401} is another relevant facet of quantum computation. The quantum state of a qubit or multiple qubits can be in a superposition state which can be further quantified by coherence. As the off-diagonal elements of a density operator represents coherence, the coherence measure based, known as $\textit{l}_{1}$-norm \cite{PhysRevLett.113.140401}, can be given by the following equation:
\begin{equation}
    C_{l1}(\rho)=\sum_{i\neq j}\rho_{ij},
\end{equation}                                             
where $\rho_{ij}$ are elements of density matrix.
\subsection{\label{sec:level2} Quantum channels}             
A quantum channel \cite{nielsen2001quantum,PhysRevA.72.062323,mishra2022attainable,utagi2020ping} is a completely positive and trace-preserving dynamical map ($\Lambda$) used to describe a quantum system's evolution in an interacting environment. The evolved density matrix of the system is given by $\Lambda(\rho)=\sum_{i}K_{i} \rho K_{i}^{\dagger}$ where $K_{i}$ represents Kraus operators- offering a simple and elegant way to invoke the system's dynamics- and satisfy the completeness relation $\sum_{i}K_{i}^{\dagger}K_{i}= I$. Such a of a dynamical map using so-called Kraus operators is called operator sum representation \cite{kraus1983states}. In this article, we have considered different noise models \cite{kumar2018non,lalita2023harnessing} and their corresponding Kraus operators as summarized in TABLE \ref{tab:noise}. \par
The efficacy of an information protocol further depends on the properties of the considered channel \cite{shadman2010optimal}. Therefore, we study the action of amplitude damping channel (ADC), random telegraph noise (RTN), power law noise (PLN), and modified Ornstein-Uhlmenbeck Noise (OUN) on the QT protocol. We also bring to the forefront the advantages of using WM and RWM in universalizing the QT protocol.

\begin{figure*}
    \subfigure[]{\includegraphics[width=0.45\linewidth]{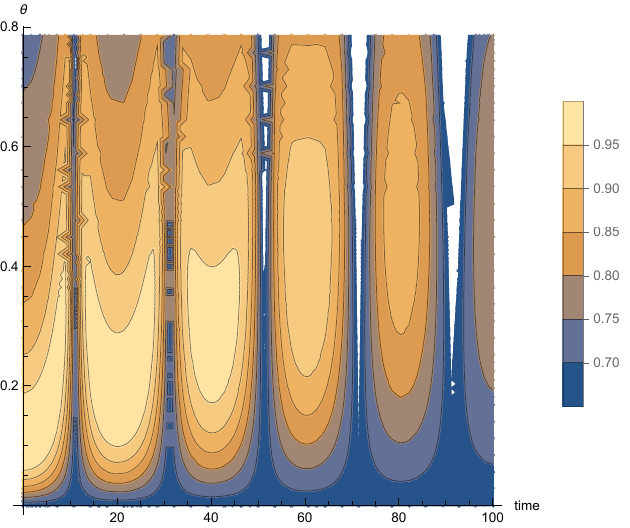}}\hfill
    \subfigure[]{\includegraphics[width=0.45\linewidth]{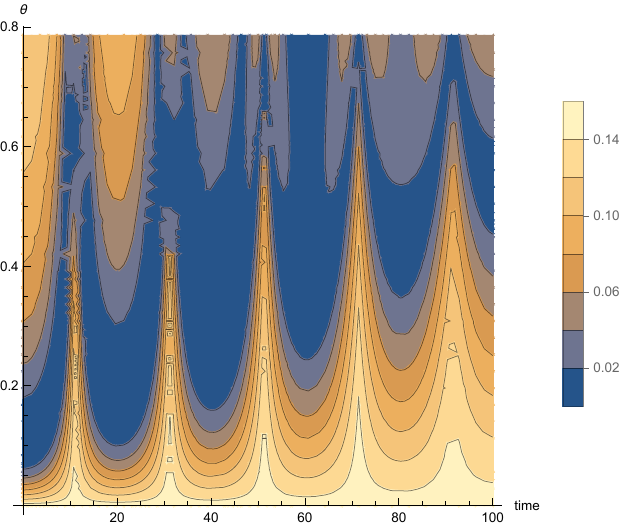}}
     \\
    \caption{ (a) WM and RWM impact on the fidelity of Bell-type state, evolving under non-Markovian AD noise ($\gamma_{0}=1$, $k=0.05$) with parameters $w=0.1$ and $wr=0.99$ (b) WM and RWM impact on the fidelity deviation of Bell-type state, evolving under non-Markovian AD noise ($\gamma_{0}=1$, $k=0.05$) with parameters $w=0.1$ and $wr=0.99$. The parameters $w$ and $wr$ are defined in Eqs. (\ref{MW}), (\ref{MWR}).}
 \label{fig: contour_plot}
\end{figure*}

\begin{figure*}
    \subfigure[]{\includegraphics[width=0.45\linewidth]{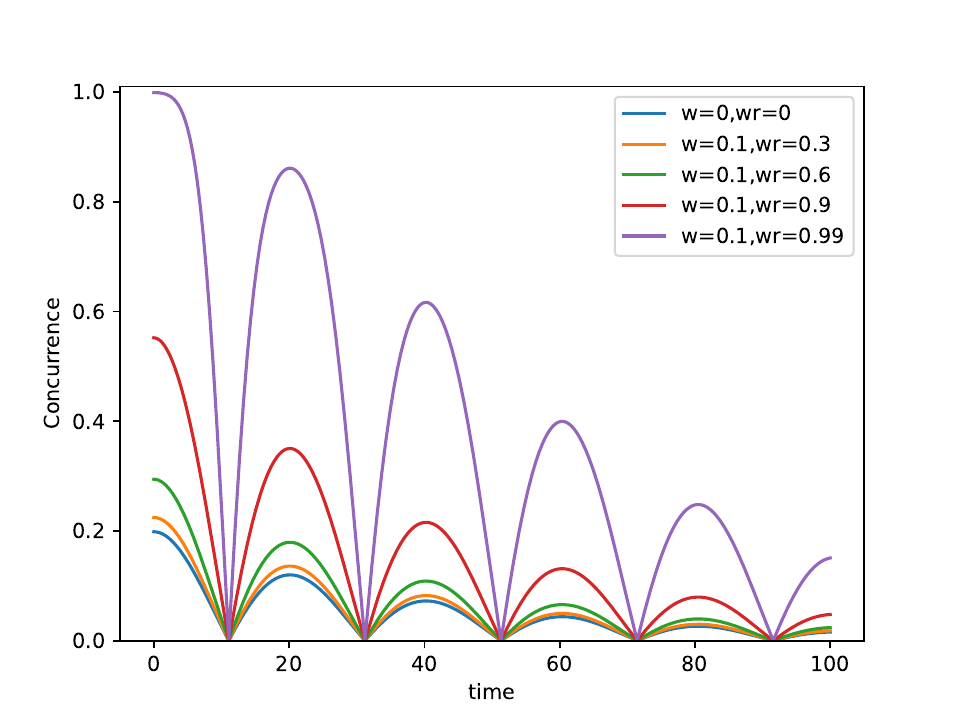}}\hfill
    \subfigure[]{\includegraphics[width=0.45\linewidth]{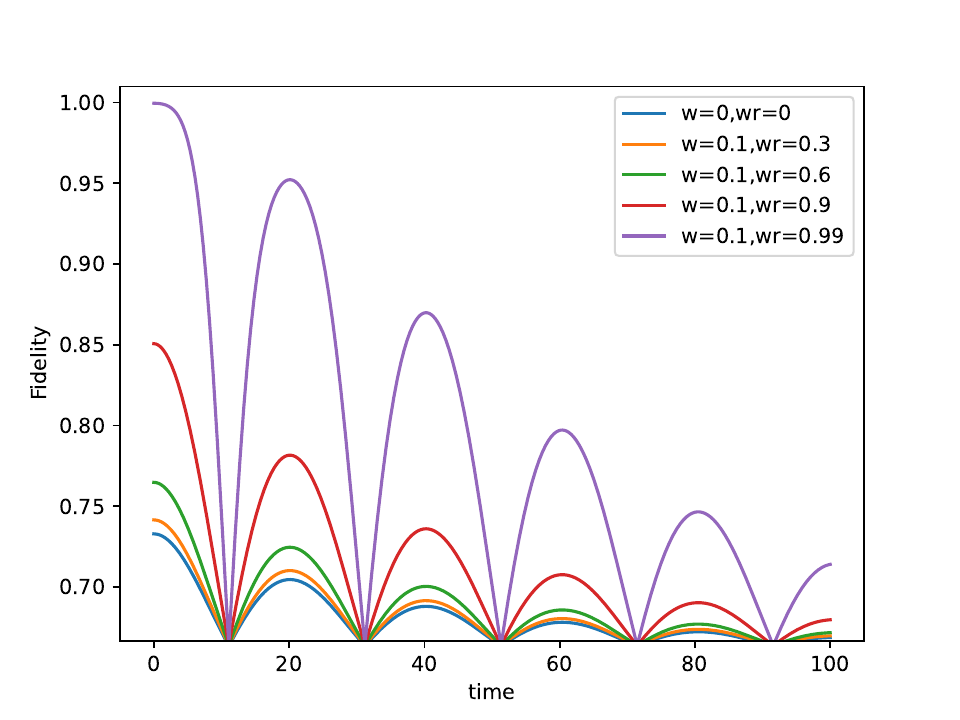}}
     \\
    \caption{WM and RWM impact on Concurrence and fidelity values of Bell-type state with Concurrence = 0.1986 evolving through non-Markovian AD noise ($\gamma_{0}=1$, $k=0.05$) (a) a concurrence of evolving state  (b) fidelity of evolving state.}
 \label{fig: WM_nadc_channel}
\end{figure*}

\subsection{ \label{sec:level3} Fidelity and Fidelity Deviations}
Fidelity is a measure of the overlap or closeness of quantum states, and therefore is very useful to analyze the success of teleportation. The average fidelity of QT is defined as
\begin{equation}
    \bigl \langle{f_{\rho}} \bigr \rangle=\int f_{\psi,\rho}  d\psi,
\end{equation}
where $ f_{\psi,\rho}=\bra{\psi} \rho \ket{\psi}$ is the overlap between input and output pair $(\ket{\psi}\bra{\psi}, \rho)$ and the integration takes place over all possible input states. Clearly, fidelity has the value one for a shared maximally entangled state. However, in the real-world scenario, available states to be used as resources are mixed states \cite{rieffel2011quantum,fano1957description} and not maximally entangled states, so achieving perfect input and output overlap is hard in reality. For a given state, the maximum average fidelity $F_{\rho}$ is a maximum value of fidelity obtained via all strategies and local unitary operations
\begin{equation}
    F_{\rho}=\max_{LU} \bigl \langle{f_{\rho}} \bigr \rangle,
\end{equation}
where optimization is performed over all possible local unitary (LU) strategies. The protocol that maximizes the fidelity values is called the optimum protocol.\par 

In addition to fidelity, fidelity deviation is another necessary quantifier of QT or UQT. It measures the deviation in fidelity for possible input states and is defined as the standard deviation of fidelity such that
\begin{equation}
    \delta_{\rho}=\sqrt{\bigl \langle{f_{\rho}^{2}}\bigr \rangle-\bigl \langle{f_{\rho}}\bigr \rangle^{2}}.
\end{equation}
Here $\bigl \langle{f_{\rho}^{2}}\bigr \rangle = \int f_{\psi,\rho}^{2}  d\psi$. 
The fidelity and fidelity deviation of a resource state for an optimum QT protocol is given by:
\begin{equation}
      F_{\rho}=\frac{1}{2}(1+\frac{1}{3}\sum_{i=1}^{3}|t_{ii}| ).
\end{equation}
 and
\begin{equation}
     \Delta_{\rho}=\frac{1}{3\sqrt{10}}\sqrt{\sum_{i< j=1}^{3}(|t_{ii}|-|t_{jj}|)^{2}},
\end{equation}
respectively. Here $t_{ii}$ are eigenvalues of the correlation matrix of the two-qubit density matrix where the $3\times3$ matrix correlation matrix ($\boldsymbol{T}$) of the two-qubit state is further given by $T_{ij}=  {\rm Tr}(\rho\sigma_{i}\otimes\sigma_{j})$; $\sigma_{i}$ indicates Pauli matrices.\par

Evidently, Bell-type states are rank-two entangled states and are not useful for \textit{universal quantum teleportation}, unless they are maximally entangled \cite{ghosal2020optimal,ghosal2020fidelity}. We bring out that non-maximally entangled Bell-type states can also be useful for UQT even in the presence of different noisy channels with the weak measurement protocol, depicted in FIG \ref{fig:wm_protocol}. We want to bring out the utility of our protocol for UQT using a non-maximally entangled state. In particular, as observed in FIG. \ref{fig: contour_plot}, for chosen weak measurement parameters (see Eqs. (\ref{MW}),~(\ref{MWR})) $w=0.1$ and $wr=0.99$, for state parameter $\theta = 0.1$, higher fidelity and lower fidelity deviation can be reached. The fidelity and fidelity deviation for given $\theta$ and time is depicted in the contour plots. The Figs. \ref{fig: contour_plot}(a) and \ref{fig: contour_plot}(b) bring out that for smaller value of $\theta$ we get less deviation in fidelity, with higher fidelity values. Interestingly, after the weak measurement protocol, the input Bell-type state with concurrence $0.1986$ evolves to an output state very close to the maximally entangled state, as seen in FIG. \ref{fig: WM_nadc_channel}(a) for a non-Markovian AD noise. The FIG. \ref{fig: WM_nadc_channel}(b) shows the variation in fidelity for given $w$ and $wr$. For the fixed value of parameter $w=0.1$, we observe enhanced fidelity as well as concurrence with increased $wr$. The Bell-type states have the generic form:
\begin{equation}
\ket{\psi}=\cos \theta \ket{00} + \sin \theta \ket{11},
\end{equation}
which coincides with Bell states for $\theta = \frac{\pi}{4}$. The maximum fidelity value for such a state is
\begin{equation}
      F_{max}=\frac{2}{3}(1+\frac{\sin{2 \theta}}{2}).
\end{equation}
Similarly, the fidelity deviation can be evaluated as 
 \begin{equation}
     \Delta=\frac{1}{3\sqrt{5}}[1-C(\theta)].
 \end{equation}
where $C(\theta)$ represents the concurrence of the two-qubit state. This shows that the state should be maximally entangled to have zero fidelity deviation \cite{Ghosal2021}. This motivates us to seek to enhance the concurrence of the state under the influence of noisy channels. To accomplish this we use WM and RWM, as they are widely used to suppress noise action. This helps to improve the quality of QT. 

\subsection{Weak Measurements and Weak Measurement Reversal}
The pre- and post-selection criteria forms the basis for weak measurement \cite{PhysRevLett.60.1351, PhysRevA.89.052105}. Since it is not a disruptive measurement, the quantum state under study can be recreated using RWM. The following non-unitary operator can be used to invoke the WM operation on a single qubit:
\begin{equation} \label{MW}
M_{w}=
\begin{pmatrix}
1 & 0 \\
0 & \sqrt{1-w} 
\end{pmatrix}.
\end{equation}
where  $0< w < 1$. The quantum state after the action of weak measurements can be recovered using RWM nonunitary operator:
\begin{equation} \label{MWR}
\qquad
\qquad
M_{wr}=
\begin{pmatrix}
\sqrt{1-wr} & 0  \\
0 & 1 
\end{pmatrix}.   
\end{equation}
where $ 0 < wr <1$. Certain protocols, such as teleportation \cite{kim2012protecting}, have benefited from using WM and RWM \cite{roy2021recycling,harraz2022enhancing,guo2020fidelity,pramanik2013improving,PhysRevLett.109.150402,hu2013non,singh2018analysing}. The experimental implementation of WM and RWM can also be performed using superconducting \cite{PhysRevLett.126.100403, PhysRevLett.101.200401} and photonic quantum systems \cite{Kim:09}. Experimental implementation of weak measurements \cite{RevModPhys.86.307} has been reported in the case of QT.  We will comment on the general behavior of weak measurements and their effect on entanglement, fidelity, and fidelity deviation in presence of correlated, Markovian and non-Markovian noises.

\section{\label{sec:level4}  Enhancing correlations and information back-flow using WM and RWM}

As a precursor to understanding the impact of non-Markovian effects on UQT using Bell-type states under WM and RWM, we study here non-Markovianity indicators like trace distance and its time derivative. The analysis begins with a Bell-type state with $\theta=0.1$. The concurrence of the state under study is equal to 0.1986. Due to less non-local correlation and concurrence, this state's ability to function well in QT is limited. We aim to understand the preservation and enhancement of non-local correlations \cite{wang2022protecting} in the present context.  This study additionally enabled us to observe the enhancement of non-Markovinity signatures. In particular, high amplification of information back-flow is observed for non-Markovian ADC noise.

\begin{figure}
         \includegraphics[width=0.45\textwidth]{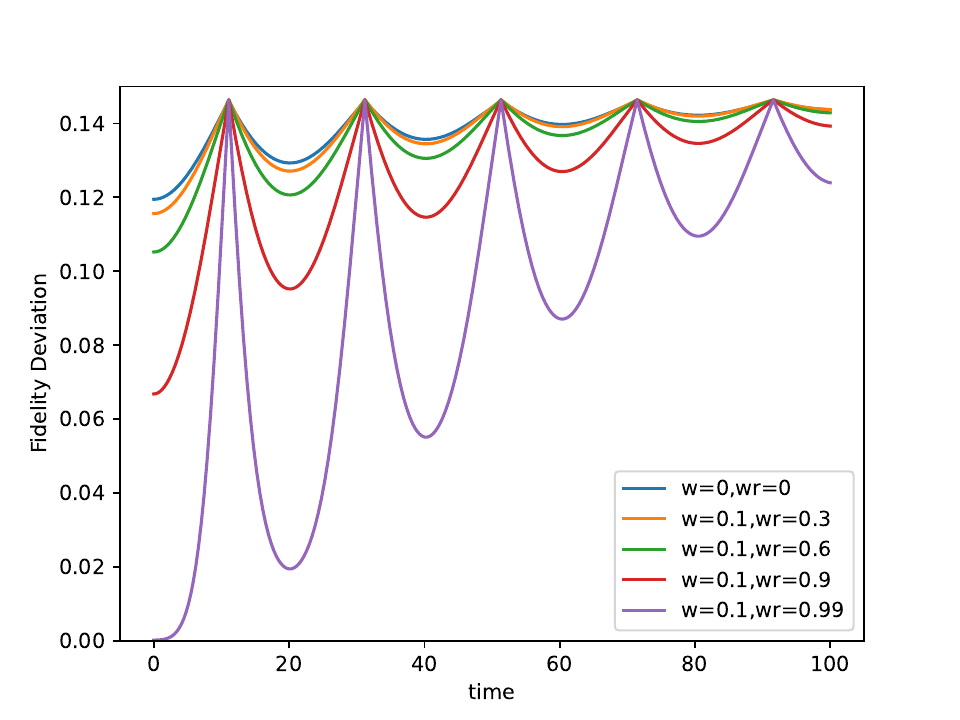}
         \caption{WM and RWM impact on fidelity deviation values of Bell-type state with Concurrence = 0.1986 evolving through non-Markovian AD noise ($\gamma_{0}=1$, $k=0.05$). }
         \label{fig:fid_dev_nadc}
\end{figure}

\subsection{\label{sec:level4i}Effect on concurrence, fidelity, and fidelity deviations}

\begin{figure*}
    \subfigure[]{\includegraphics[width=0.45\linewidth]{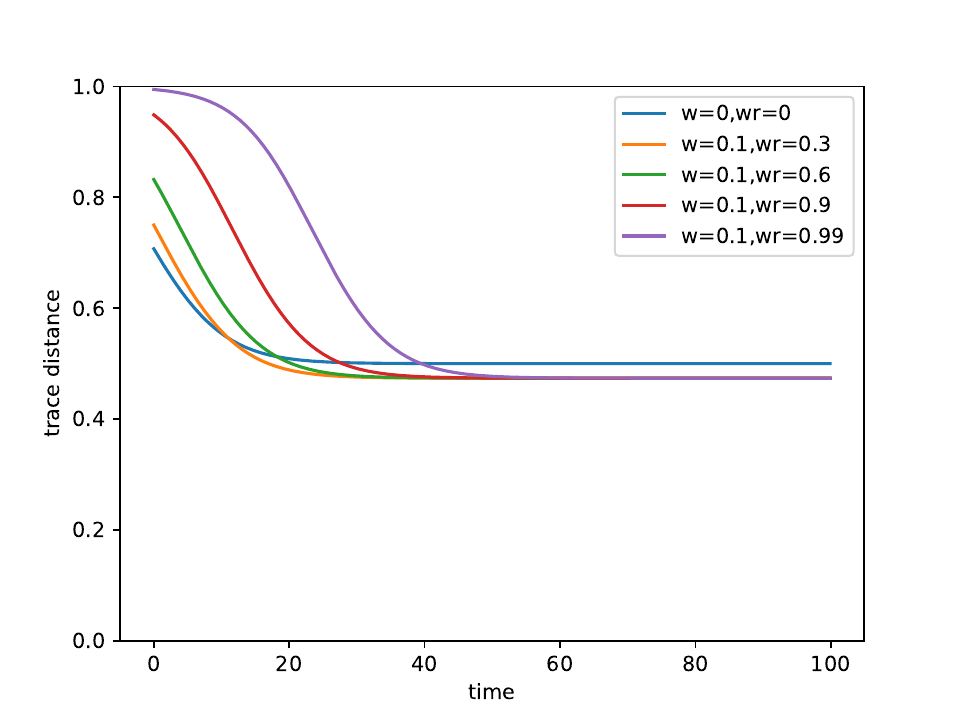}}\hfill
    \subfigure[]{\includegraphics[width=0.45\linewidth]{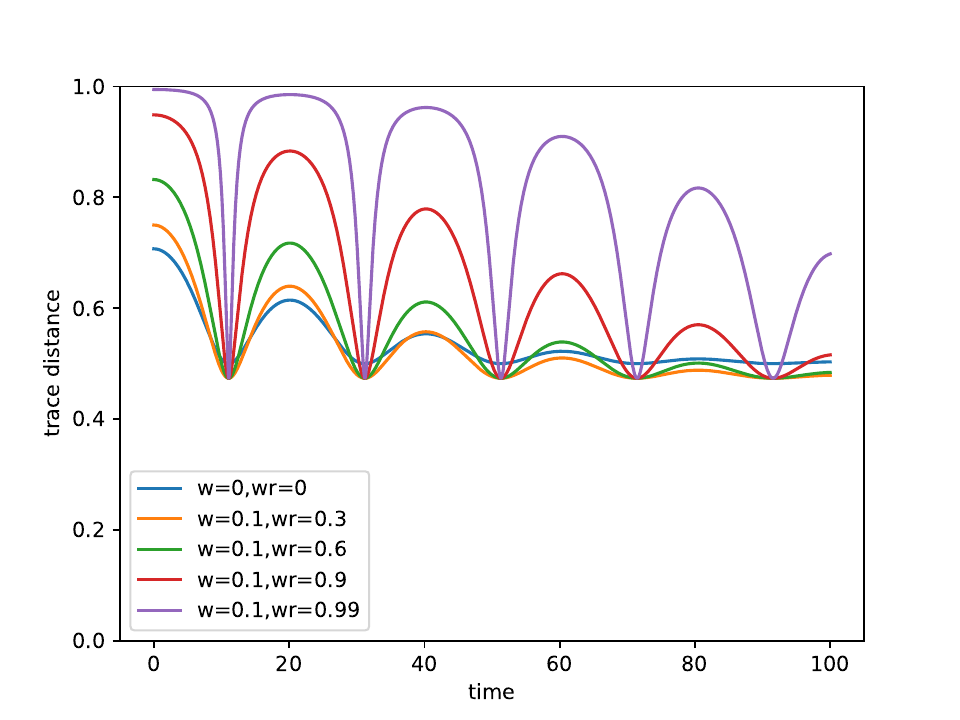}}
     \\
    \caption{WM and RWM impact on trace distance between state $\ket{00}$ and $\frac{\ket{00}+\ket{11}}{\sqrt{2}} $ (a) trace distance in ADC (markovian) ($\gamma = 0.2$) (b) trace distance in non-Markovian AD noise ($\gamma_{0}=1$, $k=0.05$). }
 \label{fig: WM_tr_madc_nmad}
\end{figure*}

In this subsection, we analyze the effect of non-Makovian AD noise on concurrence, fidelity, and fidelity deviation in the presence of weak measurements using the Bell-type state with $\theta=0.1$.
The action of WM and RWM on concurrence and fidelity in the case of evolution through non-Markovian AD noise is presented in FIG. \ref{fig: WM_nadc_channel}. The increase in concurrence and fidelity is observed for increasing RWM strength $wr$ for a fixed $w$. The increase in quantum correlation makes output state closer to the maximally entangled state after the completion of the protocol in FIG. \ref{fig:wm_protocol}.  This further reflects in the fidelity deviation values as the output state's entanglement is increased several folds after the proposed protocol. FIG \ref{fig:fid_dev_nadc} clearly demonstrates that an increase in $wr$ value causes minimization of fidelity deviations. The highest value of $wr=0.99$ allows the state to reach very less fidelity deviation for the input state with less concurrence value. In fact, the state is useful for UQT in presence of weak measurements for certain values of $t$. A detailed discussion will follow in sub-subsection \ref{subsec:control}. 

\begin{figure}
         \includegraphics[width=0.45\textwidth]{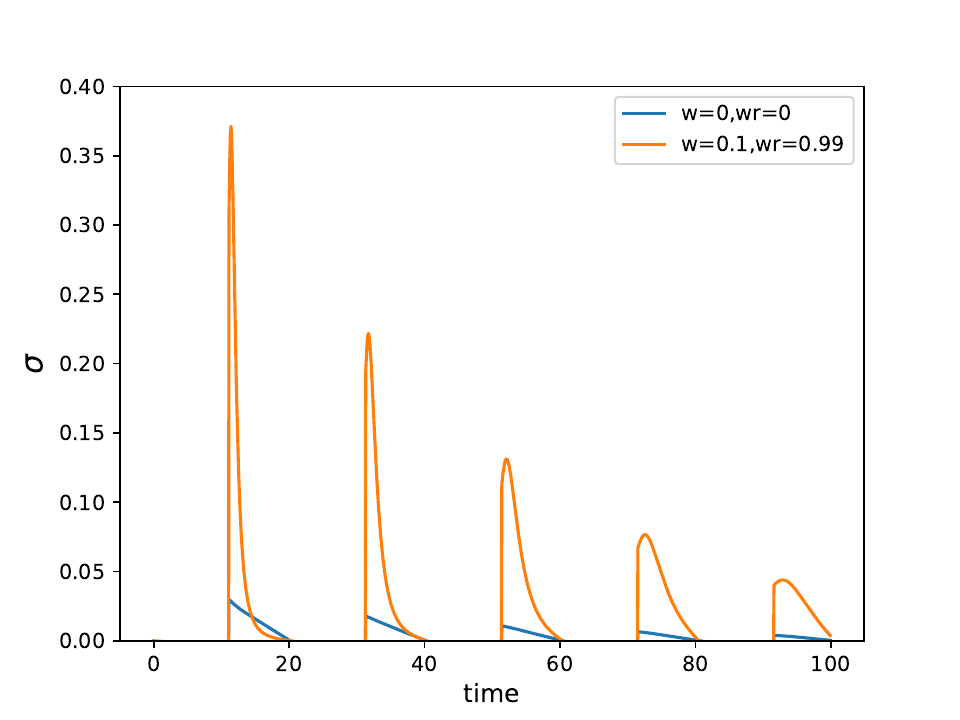}
         \caption{WM and RWM impact on BLP measure for non-Markovian AD channel  ($\gamma_{0}=1$, $k=0.05$).}
         \label{fig:BLP_Measure}
\end{figure}

\subsection{\label{sec:level4iv}WM and RWM effect on information back-flow}
Information back-flow is quantified by trace distance, which is a measure of the distinguishability of quantum states. For example, trace distance for states $\rho_{1}$ and $\rho_{2}$ is given by the following equation:
\begin{equation}
    D(\rho_{1},\rho_{2})=\frac{1}{2}Tr\abs{\rho_{1}-\rho_{2}},
\end{equation}
where $\abs{O}=\sqrt{O^{\dagger}O}$. The study of Markovian
processes shows the contracting nature of the trace distance \cite{PhysRevA.81.062115}. The contraction
in trace distance is viewed as information loss from the system to the environment. Clearly, the consequences of a system exhibiting Markovian dynamics are reflected in trace distance as a reduction in state distinguishability as a monotonic decrease in trace distance. This points towards the potential use of trace distance to study system's
dynamical evolution and information flow. Unlike the Markovian process, non-Markovian processes result in an increase in state distinguishability due to the information back-flow from environment to system \cite{motavallibashi2021non,hao2012enhanced}.
The increase in distinguishability of quantum states during evolution is beneficial as it signifies revival of quantumness of a system. In order to identify the non-Markovianity, we resort to a trace distance-based measure known as BLP (Bruer-Laine-Piilo) measure \cite{PhysRevLett.103.210401} which is the time derivative of the trace distance of two quantum states as stated below:
\begin{equation}
    \sigma (t, \rho_{1,2}(0))= \frac{\partial}{\partial t} D(\rho_{1},\rho_{2}).
\end{equation}
$\sigma > 0$ for a certain time interval implies the non-Markivianity in dynamics, in particular, and an indication of information back-flow. The action of weak measurements increases the value of $\sigma$ which in turn allows an increased information back-flow as compared to the scenario where weak measurements are not applied.

For example, we consider  the effect of weak measurement and its reversal on the dynamics of the following quantum states
\begin{eqnarray}
\ket{\psi_{1}(0)}= \ket{00},\\
\ket{\psi_{2}(0)}= \frac{\ket{00}+\ket{11}}{\sqrt{2}},  
\end{eqnarray}
evolving under the ADC channel considering both Markovian and non-Markovian regimes. FIG. \ref{fig: WM_tr_madc_nmad} depicts the same. In \ref{fig: WM_tr_madc_nmad} (a) we can see a monotonic decrease of trace distance implying Markovian nature of dynamics. FIG. \ref{fig: WM_tr_madc_nmad} (b) shows typical non-Markovian behavior because of non-monotonic nature in variation of trace distance, which enhances with increased strength of RWM. Our analysis shows the appearance of a high value of trace distance for non-Makovian dynamics, as compared to their Markovian counterpart, for the chosen value of WM and RWM parameters violating $D(\rho_{1}(t_{2}),\rho_{2}(t_{2})) \leq D(\rho_{1}(t_{1}),\rho_{2}(t_{1}))$ condition for time $t_{2} > t_{1}$. The recurring nature in FIG. \ref{fig: WM_tr_madc_nmad} (b), is a typical feature of P-indivisible nature of non-Markovian AD noise \cite{utagi2021non}. The derivative of trace distance shows a value greater than zero for certain time intervals, pointing towards an increase in distinguishability between quantum states and possible information flow back into the system. The effect of weak measurements enhances the information inflow and results in a high value to the BLP measure, as represented in FIG. \ref{fig:BLP_Measure}. The higher $wr$ values cause a greater increase in $\sigma$ values. This further suggests that control over non-Markovian features of the quantum channel can be analyzed using weak measurements. \par
The role of WM and RWM causes enhancement in non-Markovianity; this is possibly due to the modification of interaction between system-environment. This further facilitates better information backflow.

\section{\label{sec:levelWm}WM and RWM on shared bell-type resource state for QT}

\begin{figure*}
    \subfigure[]{\includegraphics[width=0.45\linewidth]{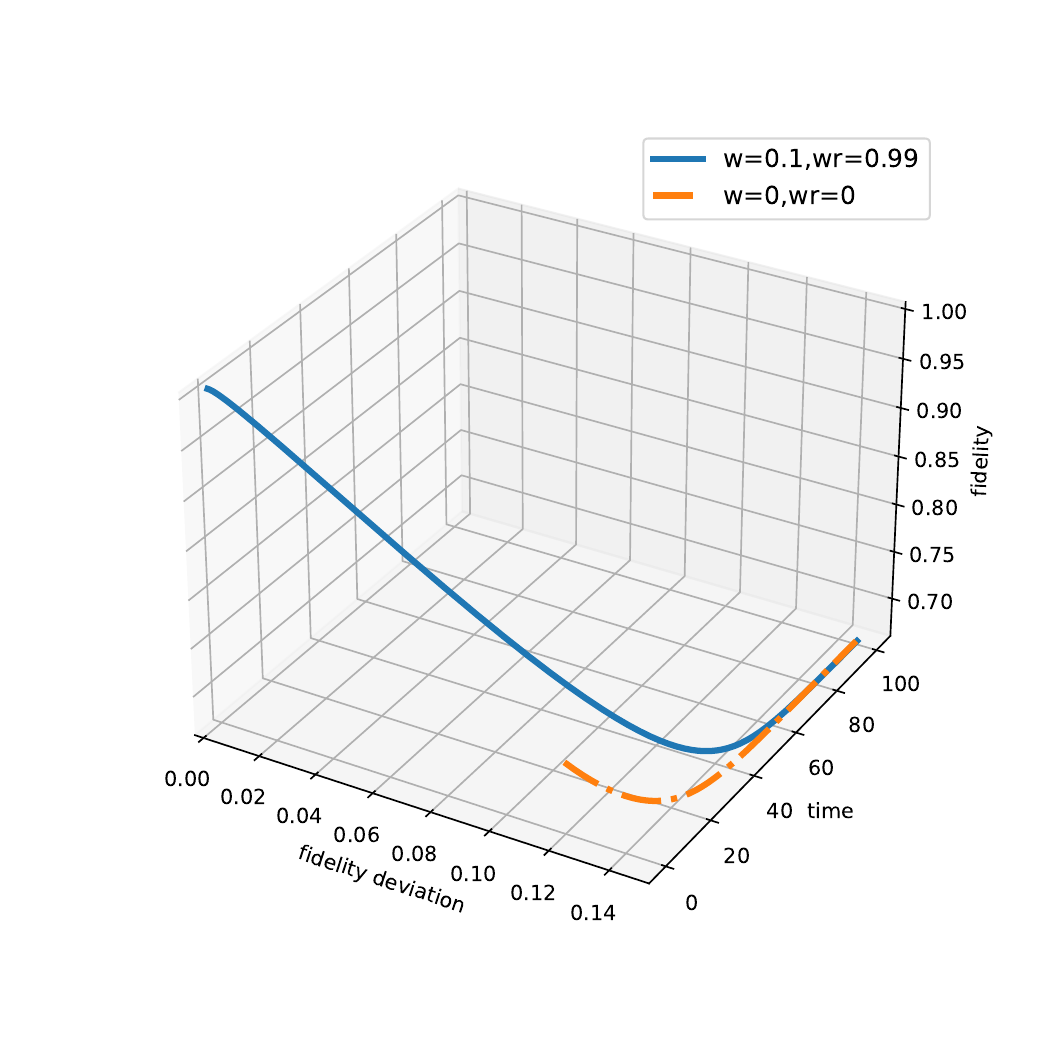}}\hfill
    \subfigure[]{\includegraphics[width=0.45\linewidth]{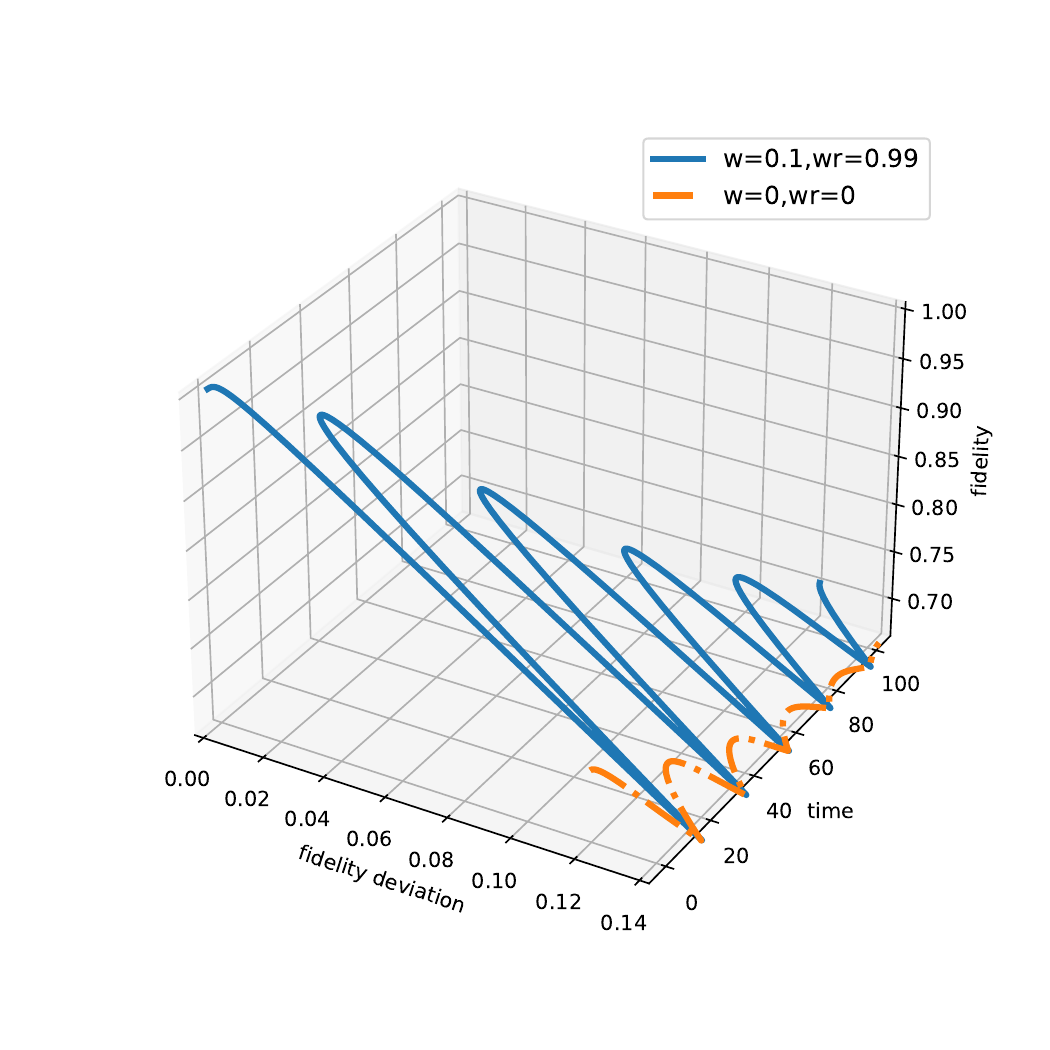}}\\
    \caption{WM and RWM impact on fidelity and fidelity deviation values of Bell-type state with concurrence = 0.1986 evolving though noisy channel  (a) ADC (Markovian) for $ \gamma =0.2$ (b) ADC (Non-Markovian) for $k=0.05, \gamma_{0} =1$. The two curves in both the figures represent the fidelity and fidelity deviation before weak measurements and after weak measurements ($w=0.1$, $wr=0.99$).}
 \label{fig: WM on fidelity and fidelity deviation ADC}
\end{figure*}

We are now in a position to discuss the impact of WM and RWM, along with non-Markovain effects, on shared Bell-type states acting as a resource for QT. Our aim would be to analyze conditions under which UQT can be achieved. The protocol from FIG. \ref{fig:wm_protocol} is looked at, where Alice prepares a two-qubit entangled quantum state, performs weak measurement on the second qubit (travel qubit), and shares it with Bob through a quantum channel. This qubit is affected by noise. The qubit that is in Alice's possession, the home qubit, is assumed to be unaffected by noise. Bob performs a weak measurement reversal operation on the received qubit before further use of the shared resource in QT. The mentioned protocol can be implemented experimentally. For example, in the case of photonic qubits, the quantum state generated can undergo WM and RWM operations using Brewster-angle glass plates \cite{Kim:09}. 

We consider Bell-type states prepared by Alice to be of the form $\ket{\psi} = \cos \theta \ket{00}+\sin(\theta) \ket{11}$ where she sends one of the qubits to Bob through the quantum channel under study. The value of concurrence of the input state to be shared is $\sin(2\theta)$, and the state becomes maximally entangled Bell state for $\theta=\frac{\pi}{4}$, reaching fidelity value of one and fidelity deviation of zero for noiseless perfect QT.  However, decoherence is inevitable as the system constantly interacts with the environment. The interaction causes a decrease in maximum fidelity and makes fidelity values deviate as time progresses. The following equation gives the dynamical map of the density matrix of a system for the protocol in FIG. \ref{fig:wm_protocol} of WM and RWM:
\begin{widetext}
\begin{equation}\label{final_state}
    \rho^f(w,wr)=\frac{(I\otimes M_{wr})\Lambda((I\otimes M_{w})\rho(I\otimes M^{\dagger}_{w}))(I \otimes M^{\dagger}_{wr}) }{Tr((I \otimes M_{wr})\Lambda((I \otimes M_{w})\rho(I \otimes M^{\dagger}_{w}))(I \otimes M^{\dagger}_{wr}))}.
\end{equation}
\end{widetext}
Here, $\Lambda(\rho)=\sum_{i=0}^{n} (I\otimes K_{i})\rho(I\otimes K_{i}^{\dagger})$ and $K_{i}$ is one of the Kraus operators corresponding to the quantum channel. The explicit form of the resulting final state for the unital and non-unital channel is given in appendix  section \ref{app:unital} and \ref{app:non-unital}.

The WM and RWM are probabilistic, and their success probability is $P_{wm}=Tr((I \otimes M_{wr})\Lambda((I \otimes M_{w})\rho(I \otimes M^{\dagger}_{w}))(I \otimes M^{\dagger}_{wr}))$. The price for high fidelity and the lowering of fidelity deviations comes with a decrease in the probability of success of the protocol $P_{wm}$. The WM and RWM can manipulate the eigenvalue of the correlation matrix $\boldsymbol{T}$ and conditionally help to satisfy the necessary relation for its eigenvalues to reach fidelity deviation zero. The optimum value of the $w$ and $wr$ parameters helps the fidelity reach higher value.\par

Similar to the non-Markovian AD noise, here we show that starting our protocol with a less entangled Bell-type state is advantageous for QT and UQT in presence of different Makovian and non-Markovian noises. The WM and RWM enhance quantum correlations, improving fidelity and minimizing the deviation in fidelity for non-maximally entangled Bell-type quantum states.

\begin{figure*}
    \subfigure[]{\includegraphics[width=0.45\linewidth]{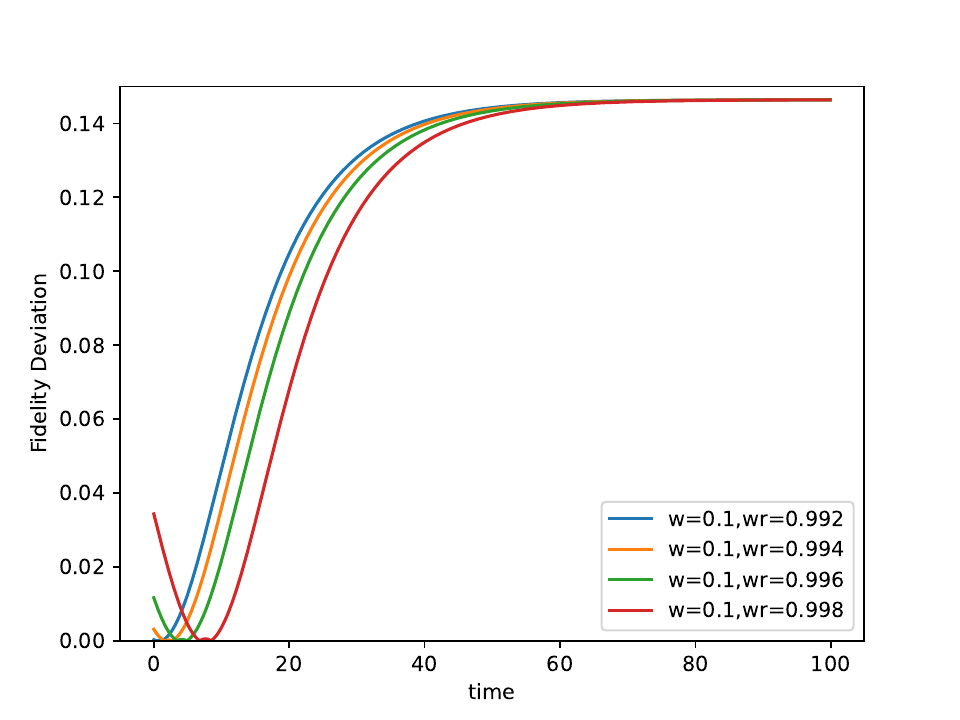}}\hfill
    \subfigure[]{\includegraphics[width=0.45\linewidth]{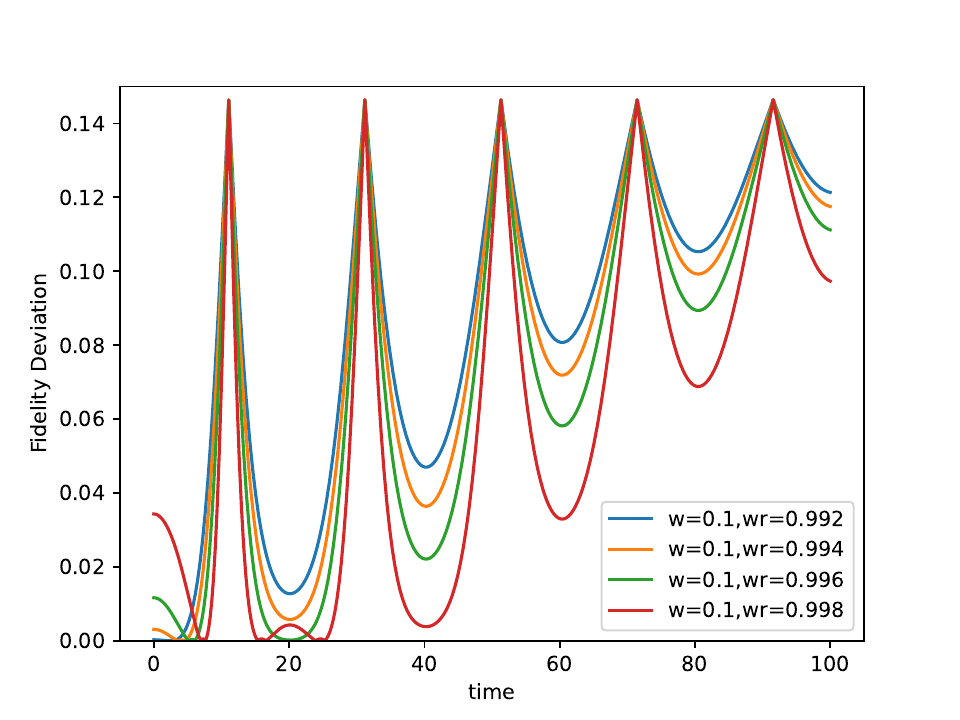}}
     \\
    \caption{(a) Weak measurement providing control over time to reach zero fidelity deviation for ADC (Markovian) ($\gamma =0.2$) with $w$ and $wr$ parameter (b) weak measurement providing control over time to reach zero fidelity deviation for non-Markovian ADC ($k=0.05,\gamma_{0} =1$) with $w$ and $wr$ parameter.}
 \label{fig:control_fid_dev}
\end{figure*}

The improvement in fidelity and fidelity deviation for a Bell-type state with a concurrence of 0.1986 is demonstrated in FIG. \ref{fig:  WM on fidelity and fidelity deviation ADC}. The input state without application of WM and RWM shows lower fidelity values and higher deviations in fidelity. The control over $w$ and $wr$ parameters helps maximize fidelity value and satisfy the minimization of fidelity deviation conditions for certain quantum channels, making it closer to achieving UQT. The non-Markovian AD channel shows very high fidelity and less fidelity deviation compared to its Markovian counterpart. Further, WM and RWM improve this scenario, bringing it closer to UQT. The parameter $w$ and $wr$ also provide control over the time at which fidelity deviation reaches zero value. This is depicted in FIG. \ref{fig:control_fid_dev}. \par

\begin{figure*}
    \subfigure[]{\includegraphics[width=0.33\linewidth]{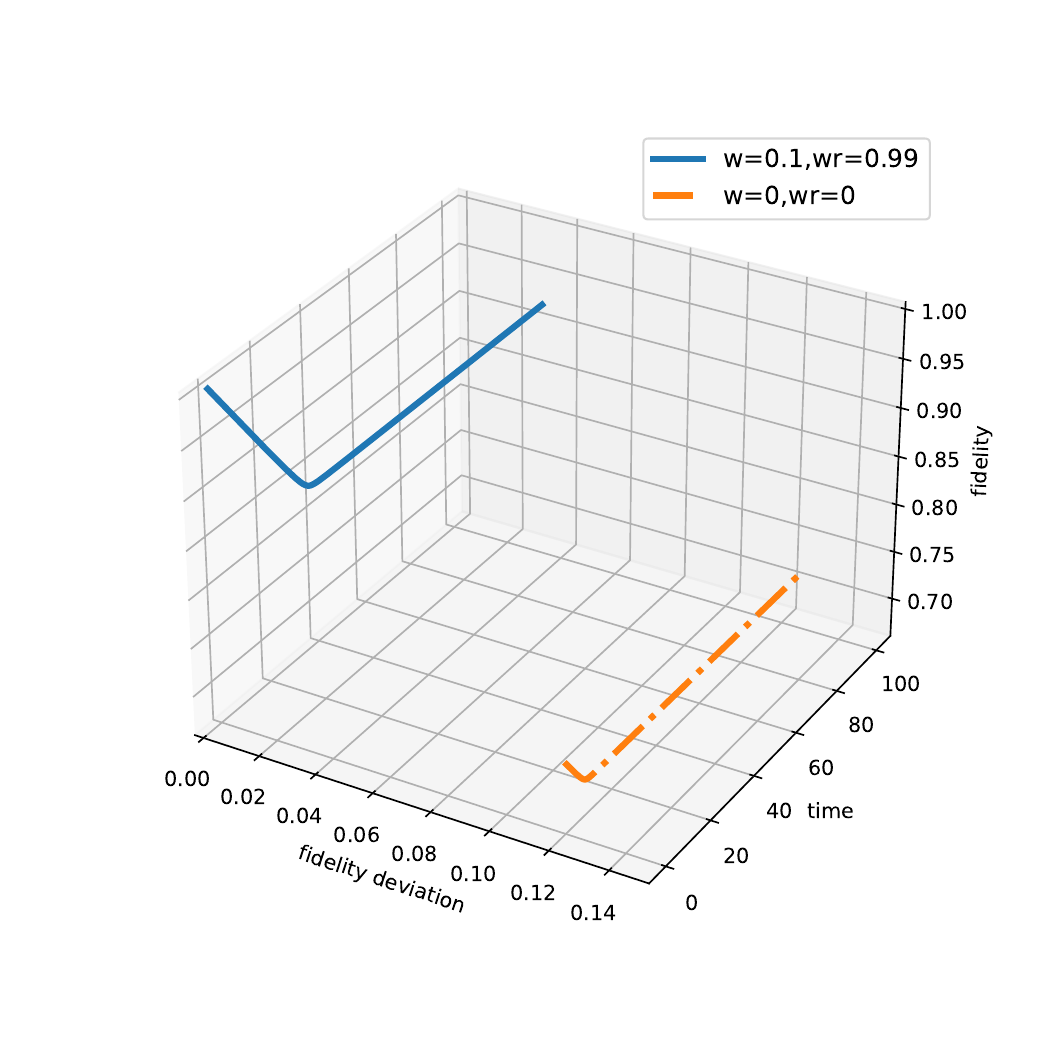}}\hfill
    \subfigure[]{\includegraphics[width=0.33\linewidth]{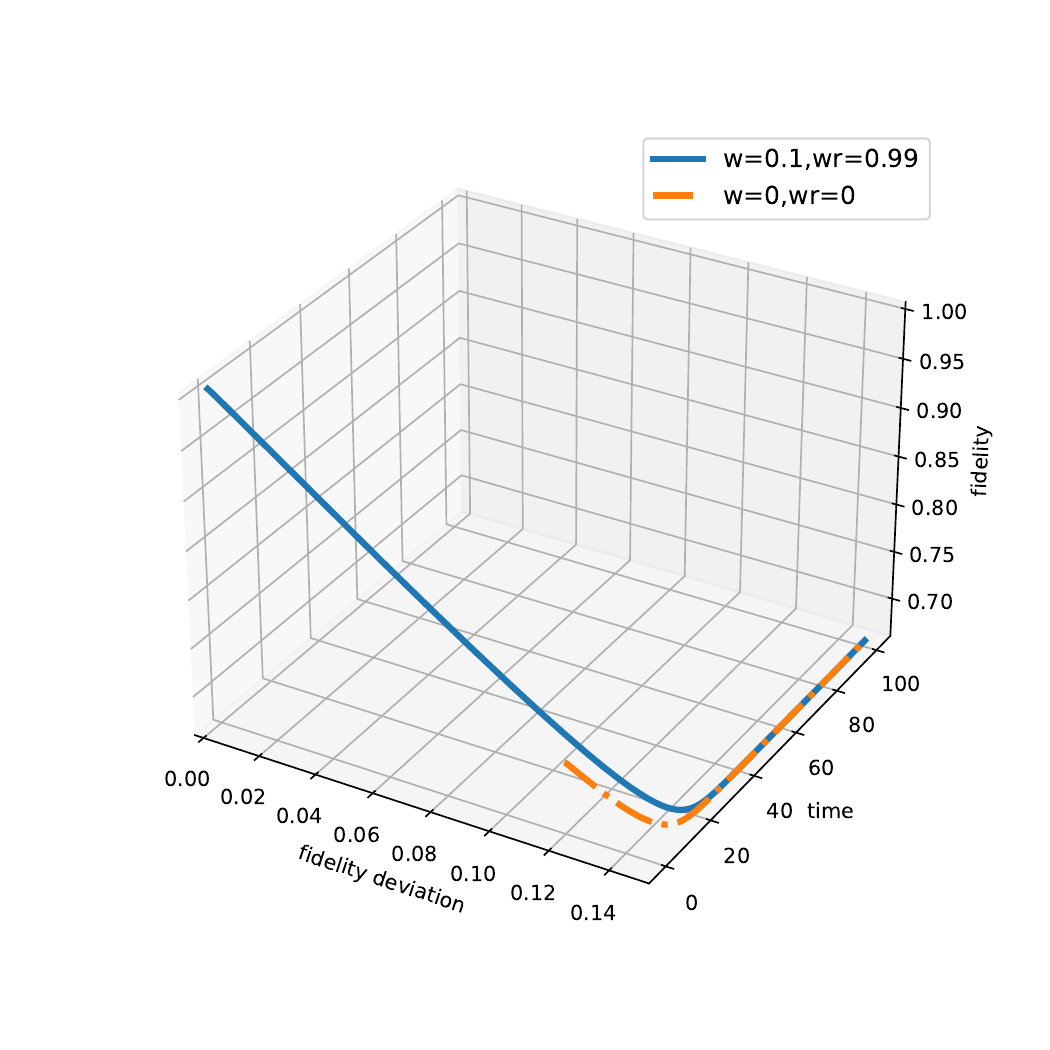}}\hfill
    \subfigure[]{\includegraphics[width=0.33\linewidth]{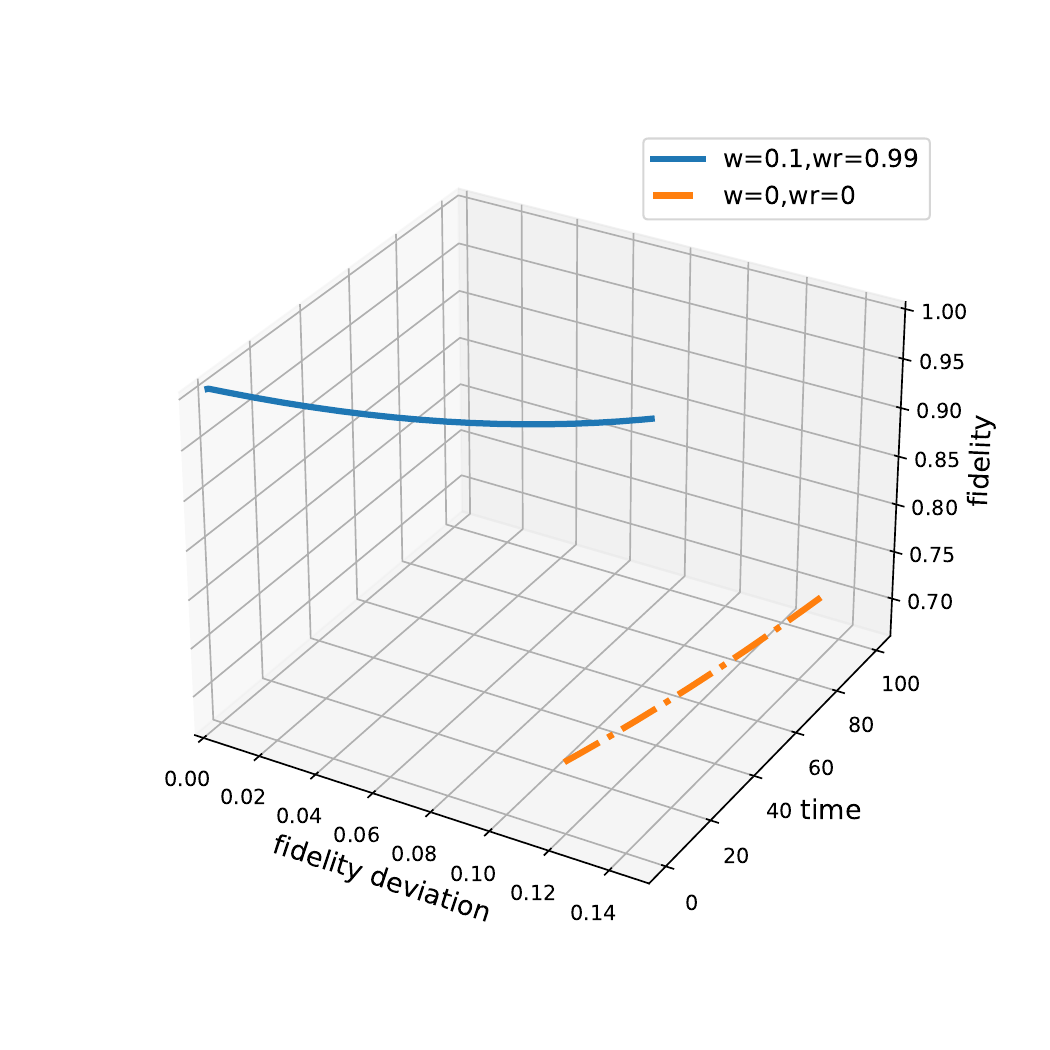}}\hfill
  
    \subfigure[]{\includegraphics[width=0.33\linewidth]{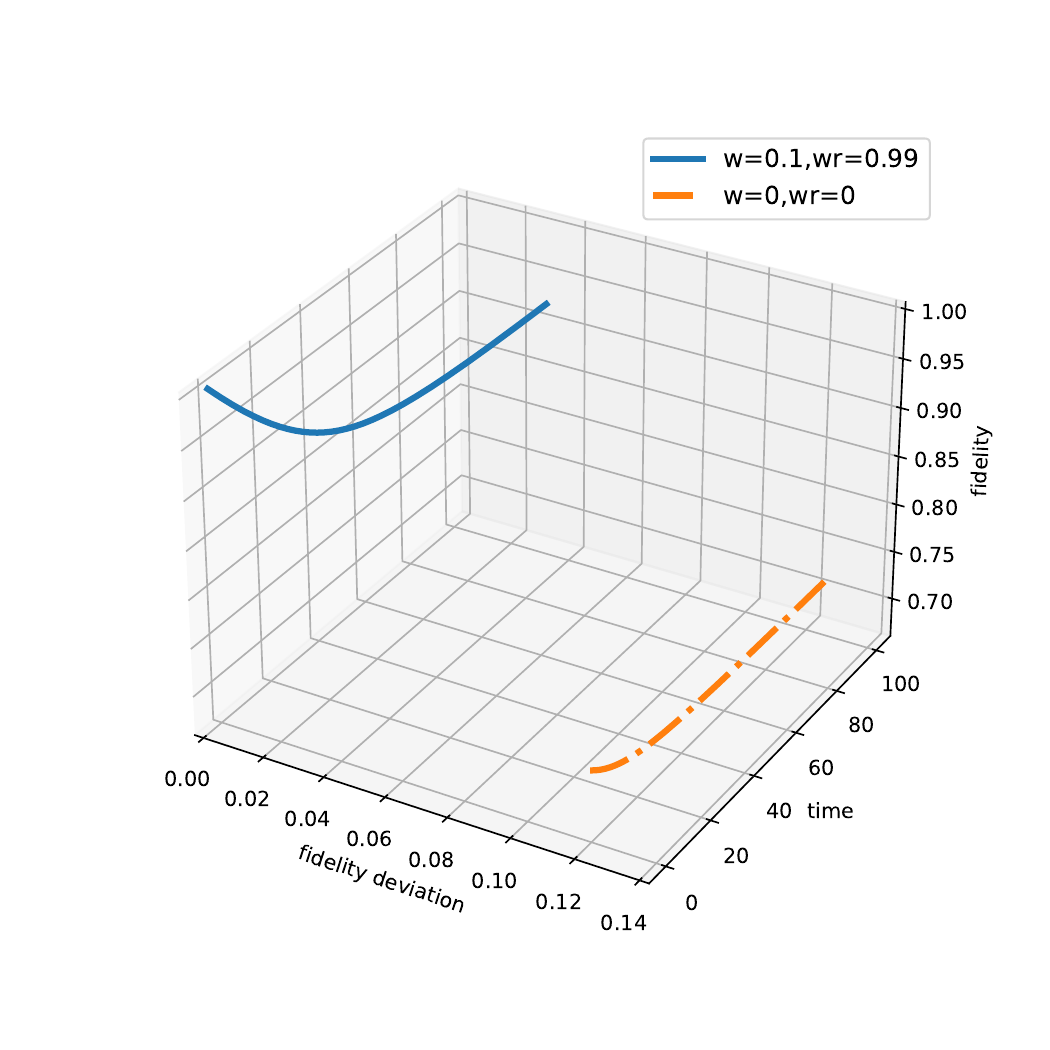}}\hfill
    \subfigure[]{\includegraphics[width=0.33\linewidth]{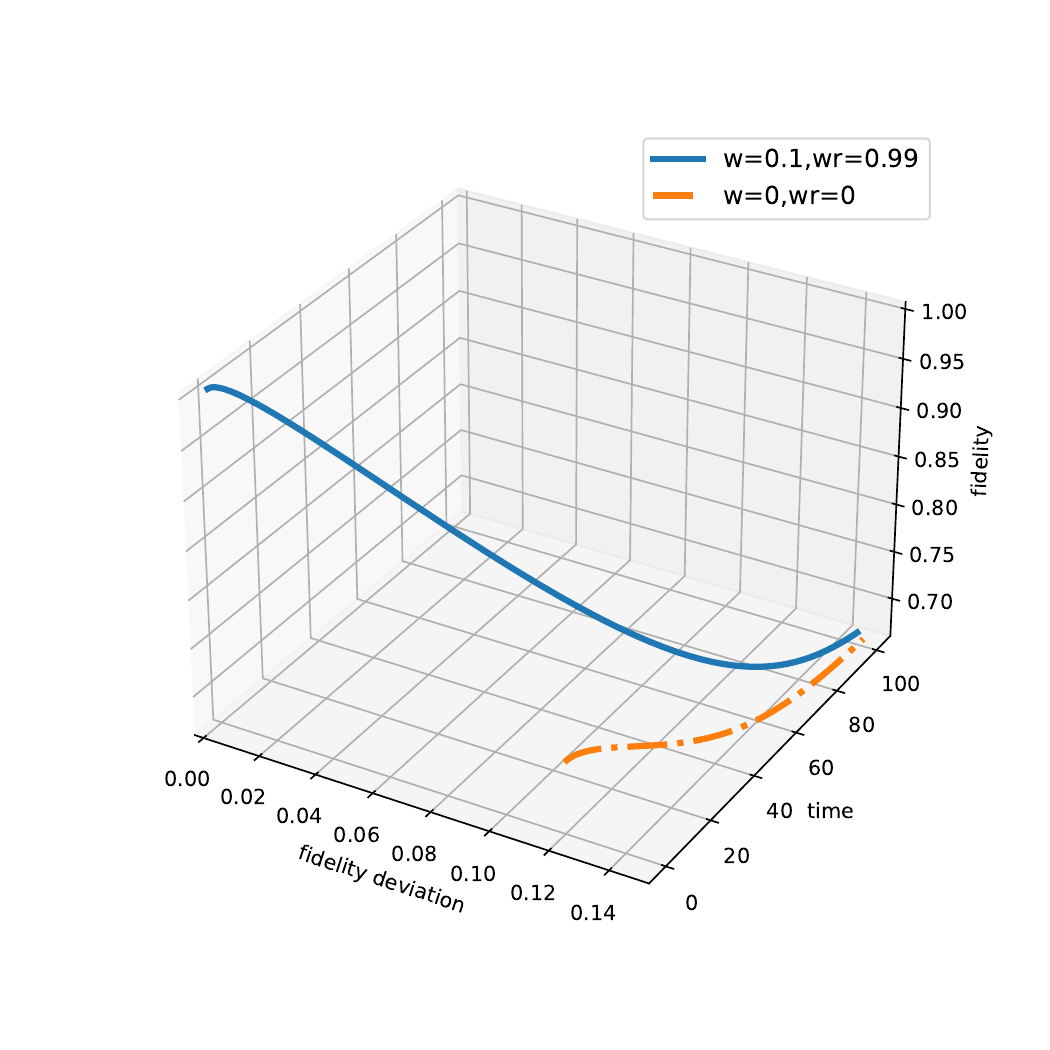}}\hfill
    \subfigure[]{\includegraphics[width=0.33\linewidth]{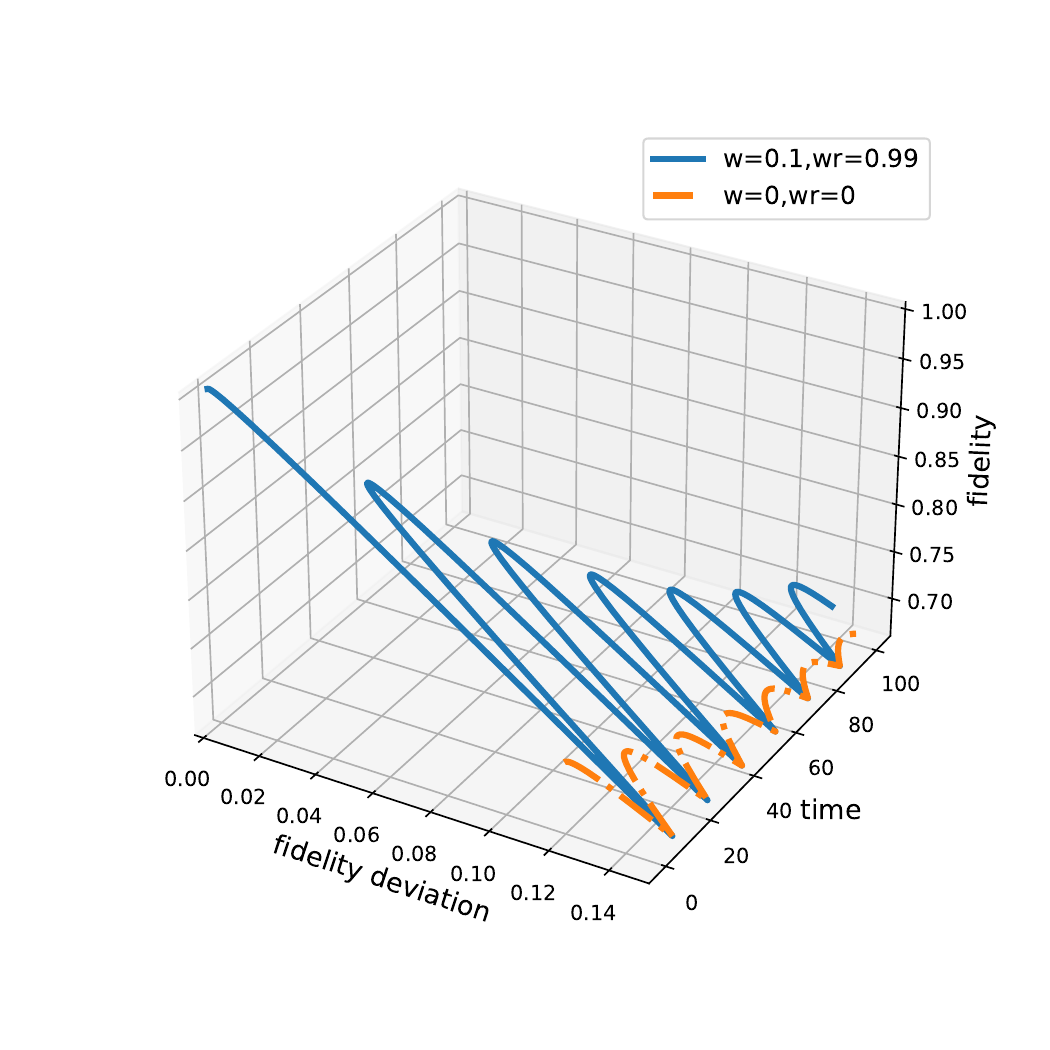}} \\
    \caption{ WM and RWM impact on fidelity and fidelity deviation values of Bell-type state with Concurrence = 0.1986 evolving though noisy channel  (a) PLN (Markovian) for $a=0.5, \gamma =3$ (b) OUN (Markovian) for $a=0.5, \gamma =5$ (c) RTN (Markovian) for $ a=0.1, \gamma =3$  (d) PLN (non-Markovian for $ a=0.5,\gamma =0.05$ (e) OUN (non-Markovian) for $a=0.1, \gamma =0.05$ (f) RTN (non-Markovian) for $a=0.1, \gamma =0.02$. The two curves in the figure represent the fidelity and fidelity deviation before weak measurements and after weak measurements ($w=0.1$,$wr=0.99$).}
\label{fig:rtn_Pln_Oun_fid_fid_dev}
\end{figure*}

The non-Markovian nature of the quantum channel is beneficial as it makes the state reach high fidelity with the necessary revival of quantum correlations. Additionally, we already demonstrated in section \ref{sec:level4} that information back-flow-related effects are visible in the output state's fidelity and fidelity deviations for a non-Markovian quantum channel. This is also true for non-Markovianity arising from the retention of correlations, which is observed for both unital PLN and OUN channels, as shown in FIG. \ref{fig:rtn_Pln_Oun_fid_fid_dev}. The information back-flow signatures disappears as the transition from non-Markovian to Markovian dynamics happens for unital RTN. In the case of PLN and OUN, the transition from non-Markovian to Markovian dynamics causes a sharp decrease in fidelity and increase in fidelity deviation.  Also, in PLN, the fidelity deviation is brought down to $ \approx 0.02$ for the presented case. This confirms the ability of WM and RWM to improve QT. \par

We also present correlated AD noise and show a direct relation between memory and correlations necessary for QT. The memory of CADC has a direct impact on fidelity and fidelity deviations as discussed in subsection IV C.

\subsection{ADC generalized expressions after weak measurements}

Our analysis of the proposed protocol in presence of ADC and weak measurements leads us to a generalized expression for concurrence. For this, we first introduce the ADC noise through operator sum representation in the Bell-type state $\rho$ using the mentioned Kraus operators in TABLE \ref{tab:noise}. The final resulting state is given by Eq. (\ref{final_state}). The probabilistic nature of WM and RWM protocol is reflected in the equation of success probability $P_{wm}$ for ADC channel given by the equation:
\begin{equation}
    (1-wr) \cos^2{(\theta)} + (1-w)(1-wr~ p(t)) \sin^2{(\theta)}.
\end{equation}
We find that the generalized expression for the final state's concurrence is: 
\begin{equation}
  \frac{\sqrt{(1-w)(1-wr)(1-p(t))}\sin{2 \theta}}{(1-wr)\cos^{2}{\theta}+(1-w)(1-wr~ p(t))\sin^{2}{\theta}}.
  \label{eq:concurrence_adc}
\end{equation}
The above expression also represents the $l_{1}Norm$ of coherence. Therefore, the concurrence, as well as coherence, is $(\sin{2\theta})$ for the input state. Both concurrence and coherence will increase many folds for the optimum values of $w$ and $wr$ for a non-maximally entangled state undergoing the protocol described, even in the presence of noise. This increase is reflected in fidelity deviations through the equation:
\begin{equation}
    \Delta=\frac{1}{3\sqrt{5}}[1-C(\theta,p(t),w,wr)].
\end{equation}
Here, $C(\theta,p(t),w,wr)$ is concurrence expressed in Eq. (\ref{eq:concurrence_adc}).
Our results show that it also helps minimize fidelity deviations at specific values of $w$ and $wr$. The action of weak measurement on fidelity and fidelity deviation for different channels is expressed in the TABLE \ref{tab:noise}, addressing the question of the possibility of zero fidelity deviation at a certain non-zero time. The results of the impact of WM and RWM on fidelity and fidelity deviation are further shown in FIG. \ref{fig: WM on fidelity and fidelity deviation ADC}.

\subsubsection{ \label{subsec:control}Control over fidelity deviation values}
WM and RWM provide control over fidelity deviation values for Bell-type states, i.e. one can control $w$ and $wr$ values to achieve zero fidelity deviation at different time. FIG. \ref{fig:control_fid_dev} represents the different values of $w$ and $wr$ for the required control to achieve zero fidelity deviation at different times in presence of both Markovian and non-Markovian AD noise. Therefore, higher $wr$ values allow information back-flow in favor of UQT. 

\subsection{RTN, PLN and OUN noise generalized expressions after weak measurements}
We now proceed to study effects of weak measurements in presence of unital RTN, PLN, and OUN noise \cite{kumar2018non}. These noises are interesting to study as they have distinct regions of Markovianity and non-Markovianity. The change in the noise parameters can cause noise to function as Markovian or non-Markovian. In this case, the success probability can be represented as:
\begin{equation}
   P_{wm} = (1-wr) \cos^{2}{ \theta } + (wr-w) \sin^{2}{ \theta }.
\end{equation}  

A further analysis provides the generalized expression for concurrence after the application of WM and RWM, and is exactly equal to the $l_{1}Norm$ of coherence, namely:
\begin{equation}
     \frac{2\sqrt{(1-w)(1-wr)}\sin{2\theta}p(t)}{(1-w)+(1-wr)-(wr-w)\cos{2\theta}}.
\end{equation}

The impact of WM and RWM on fidelity and fidelity deviation is presented in FIG. \ref{fig:rtn_Pln_Oun_fid_fid_dev}. At the mentioned values of $w$ and $wr$, there is a high improvement in fidelity as well as fidelity deviation. The memory effects from non-Markovian noise also cause enhancement in results because of the discussed increase in information back-flow as well as retention of correlation for a longer time. 

\begin{table*}
\caption{\label{tab:noise}Different Noise and possibility of reaching fidelity deviation values zero in studied protocol
}
\begin{ruledtabular}
\begin{tabular}{cccccccc}
$Noise$ & $Kraus Operators$ &$Noise Functon$ &$\Delta(w,wr)=0  (t > 0)?$     \\ \hline
& & & & \\
ADC(Markovian) &$
K_{0}=
\begin{pmatrix}
1 & 0\\
0 & \sqrt{1-p(t)} 
\end{pmatrix} ,
K_{1}=
\begin{pmatrix}
0 & \sqrt{p(t)}\\
0 & 0 
\end{pmatrix}    
$&$p(t)=1-\exp[-\gamma t]$& Yes\\
& & & & \\

OUN(Markovian)&$K_{0}=\sqrt{\frac{1+p(t)}{2}}\boldsymbol{I}, \sqrt{\frac{1-p(t)}{2}}\sigma_{z}$& $ p(t)=\exp[-\frac{a}{2}(t+\frac{1}{\gamma}(\exp[-\gamma t]-1))]$ &No\\
& & & & \\
PLN(Markovian)&$K_{0}=\sqrt{\frac{1+p(t)}{2}}\boldsymbol{I}, \sqrt{\frac{1-p(t)}{2}}\sigma_{z}$&$p(t)=\exp[-\frac{t(t\gamma+2)a \gamma}{2(t \gamma +1)^2}]$ &No\\
& & & & \\

RTN(Markovian)&$K_{0}=\sqrt{\frac{1+p(t)}{2}}\boldsymbol{I}, \sqrt{\frac{1-p(t)}{2}}\sigma_{z}$& $p(t)=\exp[-\gamma t](\cos{(\omega \gamma t)}+\frac{\sin{(\omega \gamma t)}}{\omega}) $ &No\\
& &$\omega= \sqrt{(\frac{2a}{\gamma})^{2}-1}$ & & \\
& & & & \\
ADC(non-Markovian)&$K_{0}=
\begin{pmatrix}
1 & 0\\
0 & \sqrt{1-p(t)} 
\end{pmatrix} ,
K_{1}=
\begin{pmatrix}
0 & \sqrt{p(t)}\\
0 & 0 
\end{pmatrix}  $& $p(t)=1-\exp[-\gamma t](\cos{(\frac{lt}{2})}+\frac{k}{l}\sin{(\frac{lt}{2})})^2$& Yes\\
 & & $l= \sqrt{2 \gamma_{0} k - k^{2}}$  & \\
 & & & & \\

RTN(non-Markovian)&$K_{0}=\sqrt{\frac{1+p(t)}{2}}\boldsymbol{I}, \sqrt{\frac{1-p(t)}{2}}\sigma_{z}$& $p(t)=\exp[-\gamma t](\cos{(\omega \gamma t)}+\frac{\sin{(\omega \gamma t)}}{\omega}) $ &No\\
& & $\omega= \sqrt{(\frac{2a}{\gamma})^{2}-1}$ &  & \\
& & & & \\
OUN(non-Markovian)&$K_{0}=\sqrt{\frac{1+p(t)}{2}}\boldsymbol{I}, \sqrt{\frac{1-p(t)}{2}}\sigma_{z}$& $ p(t)=\exp[-\frac{a}{2}(t+\frac{1}{\gamma}(\exp[-\gamma t]-1))]$&No\\
& & & & \\
PLN(non-Markovian)&$K_{0}=\sqrt{\frac{1+p(t)}{2}}\boldsymbol{I}, \sqrt{\frac{1-p(t)}{2}}\sigma_{z}$& $p(t)=\exp[-\frac{t(t\gamma+2)a \gamma}{2(t \gamma +1)^2}]$ & No\\
& & & & \\
Correlated ADC & $E_{0}= {\rm Eq.} ~(\ref{subeq:k3}), E_{1}= {\rm Eq.}~(\ref{subeq:k4})$ &$p(t)=1-\exp[-\gamma t]$& Yes &
\end{tabular}
\end{ruledtabular}
\end{table*}

\begin{figure*}
    \subfigure[]{\includegraphics[width=0.45\linewidth]{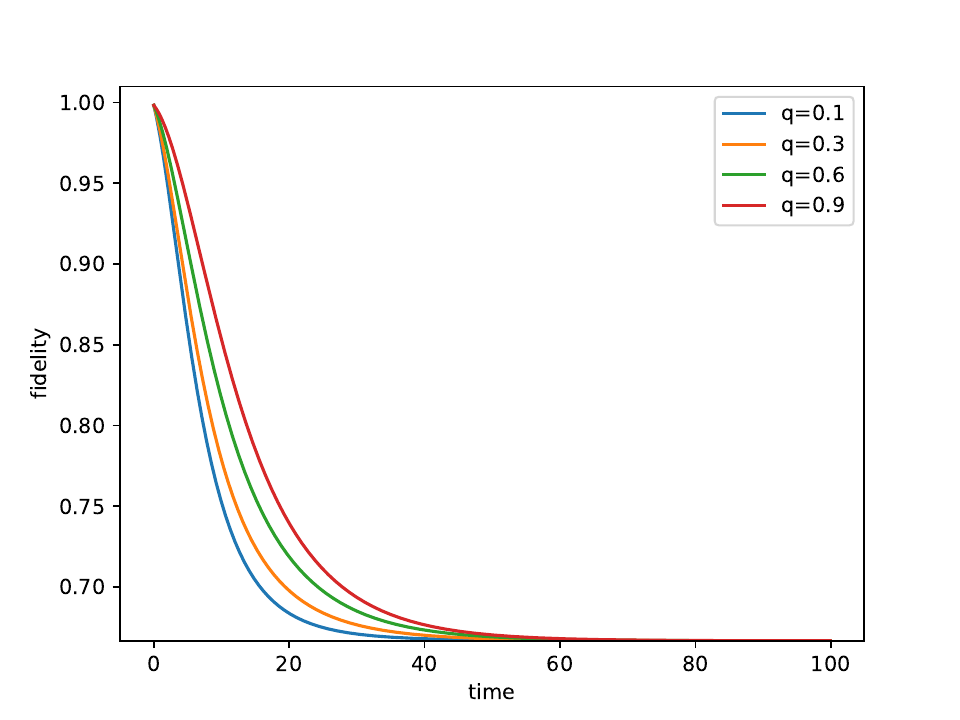}}\hfill
    \subfigure[]{\includegraphics[width=0.45\linewidth]{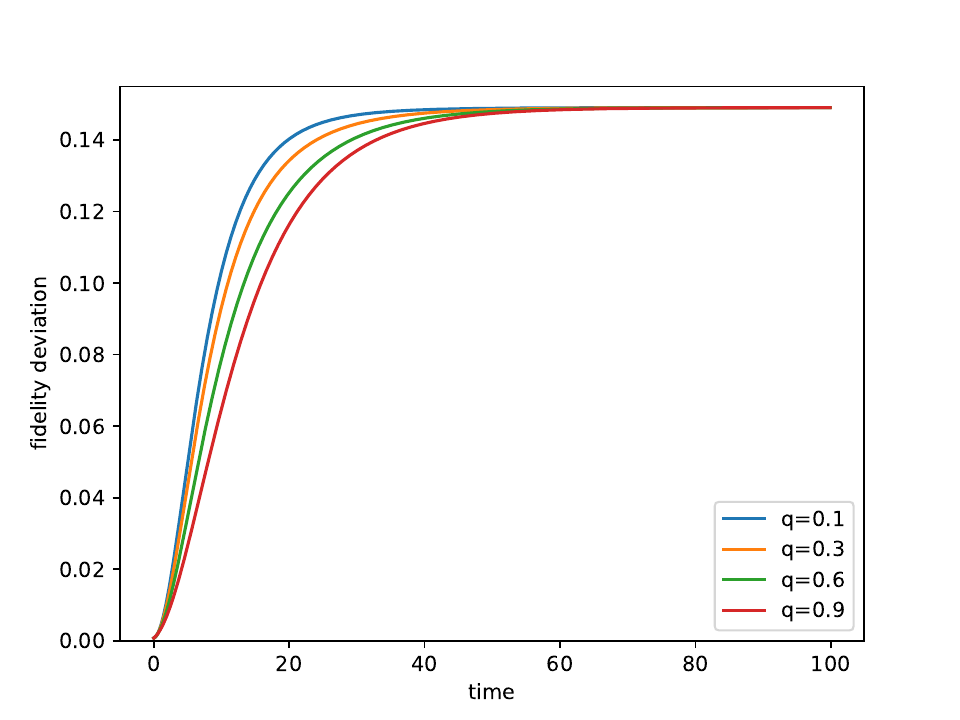}}
     \\
    \caption{(a) The memory factor q impact on the fidelity in CADC ($\gamma=0.2$)  with $w=0.1$ and $wr=0.9$ parameters (b) memory factor impact on the fidelity and fidelity deviation in CADC ($\gamma=0.2$) with $w=0.1$ and $wr=0.9$.}
 \label{fig: WM_cadc_memory}
\end{figure*}

\subsection{WM and RWM protocol on shared Bell-type state through correlated amplitude damping channels}

\begin{figure*}
    \subfigure[]{\includegraphics[width=0.45\linewidth]{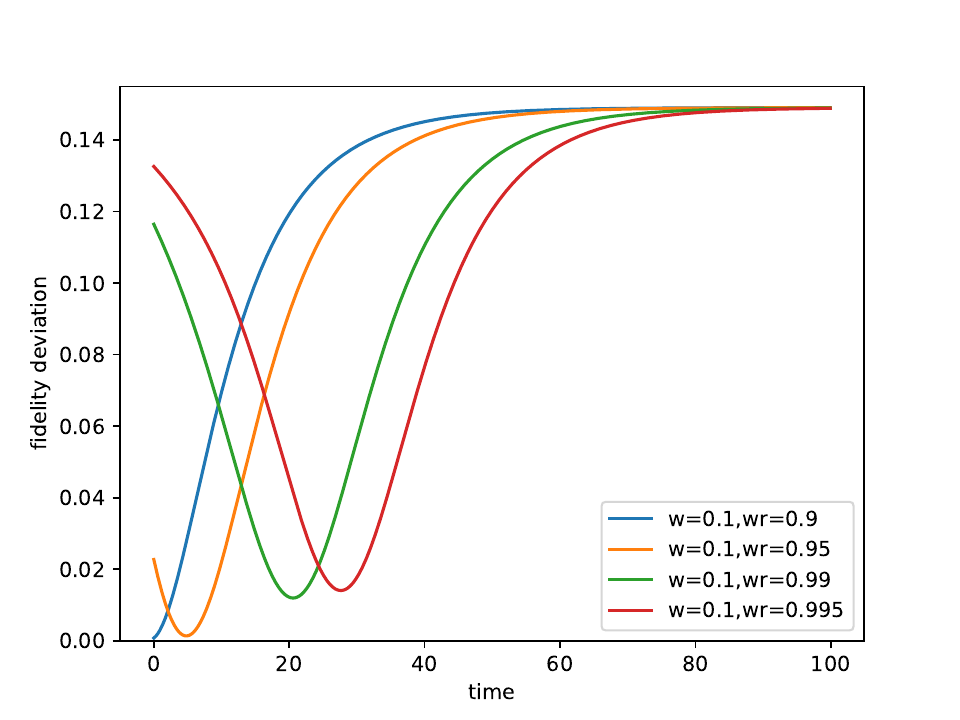}}\hfill
    \subfigure[]{\includegraphics[width=0.45\linewidth]{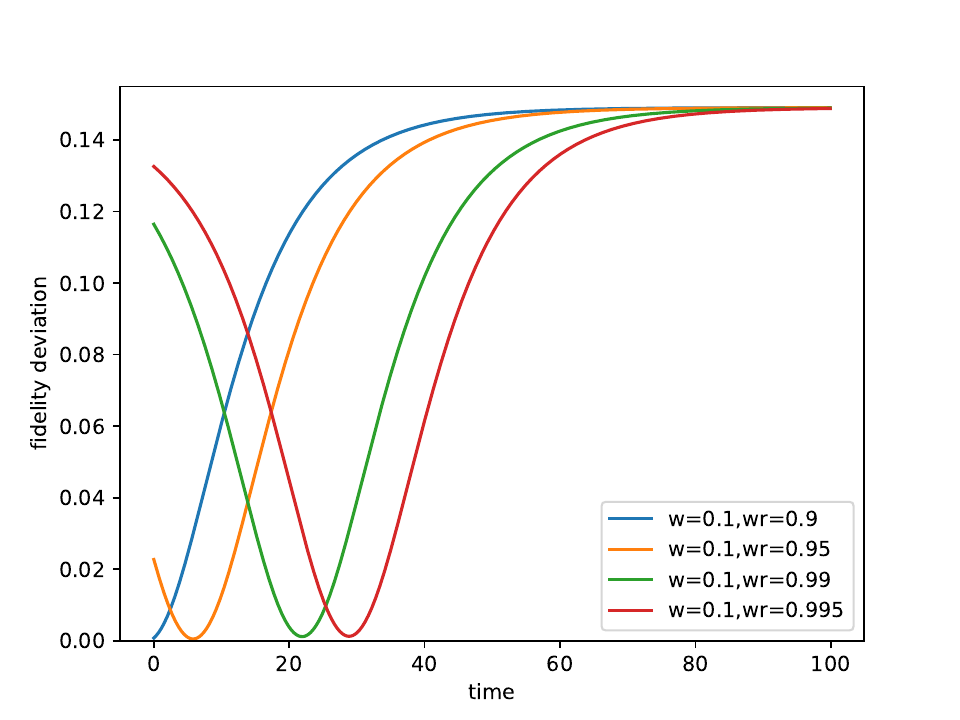}}
     \\
    \caption{(a) Fidelity deviation in CADC ($\gamma=0.2$, q=0.8) with $w$ and $wr$ parameters (b) fidelity deviation in CADC ($\gamma=0.2$, q=0.99) with $w$ and $wr$ parameters.}
 \label{fig: WM_cadc_memory_control}
\end{figure*}

The effect of memory is a necessary part that needs attention as it causes enhancement in quantum correlations for the above-studied noise models \cite{kumar2018non}. In order to analyze the impact of memory factor in presence of weak measurements, we further study the correlated amplitude damping channel (CADC) \cite{xiao2016protecting,PhysRevA.67.064301}. For this, the protocol in FIG. \ref{fig:wm_protocol} is modified to allow the action of WM and RWM on the second qubit as well considering that CADC involves noise impacting both qubits. The protocol is as follows:\\
Alice prepares two-qubit entangled quantum states, performs WM on both qubits and sends one qubit each to both Bob and Charlie through CADC channels. The qubits shared to Bob and Charlie are undergoing the noisy evolution through CADC. Both Bob and Charlie perform  RWM on their respective qubits upon receipt. The resulting final state is evaluated for UQT. Here, CADC involves Kraus operators which cannot be expressed as tensor products of individual single-qubit Kraus operators and hence the memory or correlated ADC. \par

The correlated amplitude damping channel is expressed as:
\begin{equation}
     \rho(t)= (1-q)\Lambda_{uncorr}(t)(\rho) + q\Lambda_{fcorr}(t)(\rho ),  
     \label{eq:cadc_corr}
\end{equation}
where the subscript $uncorr$ represents a two-qubit uncorrelated AD noise and $fcorr$ represents a two-qubit fully-correlated AD noise; q being the memory factor such that $0 \leq q \leq 1$. 
The uncorrelated part in Eq. \ref{eq:cadc_corr} is expressed using the following equation:
\begin{equation}
     \Lambda_{uncorr}(t)(\rho)= \sum_{i,j}(K_{i} \otimes K_{j}) \rho (K_{i}^{\dagger} \otimes K_{j}^{\dagger}).
\end{equation}
where $K_{i}$ are Kraus operators of ADC (Markovian) such that:

\begin{subequations}
\label{eq:whole1}
\begin{eqnarray}
    K_{0}=\begin{pmatrix}
    1 & 0\\
    0 & \sqrt{1-p(t)} 
    \end{pmatrix} ,\label{subeq:a3}
\\\
    K_{1}=
    \begin{pmatrix}
    0 & \sqrt{p(t)}\\
    0 & 0 
    \end{pmatrix},\label{subeq:a4}
\end{eqnarray}
 \label{eq:AD}
\end{subequations}
where $p(t)=1-\exp[-\gamma t]$.
    
Similarly, the correlated part in Eq. \ref{eq:cadc_corr} is expressed using the following equation:
\begin{equation}
     \Lambda_{fcorr}(t)(\rho) = \sum_{i}E_{i} \rho E_{i}^{\dagger}.
\end{equation}
where, $E_{i}$ are two-qubit Kraus operators as expressed in Eq. (\ref{eq:CAD}). The Kraus operators used to represent $fcorr$ channel allow for the introduction of memory. The uncorrelated part of the quantum channel involves a tensor product of the map, implying a series of
processes happening on the qubits without any correlation. For the correlated part of the channel, channel evolution correlates the noise effects. Hence, the transformation on the second qubit depends on the
first one. This is kind of memory effect where previous history impacts later dynamics. Thus, the factor $q$ is called
the memory factor, which controls the memory effects. We study the effect of memory on fidelity and fidelity deviation in presence of CADC by varying the memory factor \textbf{q}. FIG. \ref{fig: WM_cadc_memory} clearly depicts the impact of memory factor on fidelity and fidelity deviation considering $w=0.1$ and $wr=0.9$. Therefore, our analysis shows that memory effects contribute positively to UQT by minimizing the fidelity deviation.

\begin{subequations}
\label{eq:whole}
\begin{eqnarray}
E_{0}=
\begin{pmatrix}
1 & 0 & 0 & 0\\
0 & 1 & 0 & 0\\
0 & 0 & 1 & 0\\
0 & 0 & 0 & \sqrt{1-p(t)}
\end{pmatrix} ,\label{subeq:k3}
\\\
E_{1}=
\begin{pmatrix}
0 & 0 & 0 & \sqrt{p(t)}\\
0 & 0 & 0 & 0\\
0 & 0 & 0 & 0\\
0 & 0 & 0 & 0
\end{pmatrix} .\label{subeq:k4}
\end{eqnarray}
 \label{eq:CAD}
\end{subequations}

Furthermore, FIG. \ref{fig: WM_cadc_memory_control} represents the effects of memory for a fixed $w$ with varying $wr$. Clearly, for a CADC close to a fully-correlated AD channel, one can achieve zero fidelity deviation even at non-zero $t$. In the absence of memory, it is not possible to reach zero fidelity deviation even with the aid of WM and RWM. \par

The presented calculations were done using the numerical simulation tool for open quantum systems called QuTip \cite{JOHANSSON20121760,JOHANSSON20131234}.

\section{\label{sec:level6} Conclusion}
The applications of weak measurements in restoring entanglement and nonlocality with improved teleportation fidelity motivated us to analyze the fidelity deviation and role of weak measurement (WM) and reversal of weak measurement (RWM) in realizing UQT. Further, in order to understand the role of non-Markovian behavior in this process, we took up different noise models in which one can make a transition from Markovian to non-Markovian regime and \textit{vice-versa}, by tuning the relevant parameters of the model. The non-Markovian effects arising either from the retention of correlations or information back-flow are found to be significantly beneficial for maximizing fidelity and minimizing fidelity deviation. The chosen noise models belonged to both unital as well as non-unital classes. Our analysis showed the use of WM and RWM in minimizing fidelity deviations for non-maximally entangled Bell-type states.  It also enhances the coherence of the state under consideration. The parameters $w$ and $wr$ provide control over achieving optimum eigenvalues of the correlation matrix (T), by which one can have minimum fidelity deviations. For the ADC and RTN channels, the impact of weak measurement on trace distance of the protocol's output states is studied. A significant increase in the non-Markovian measure is observed using weak measurements. \par 

Endeavor is made to understand the memory effects and their impact on fidelity deviation for the correlated Amplitude Damping channel (CADC). The magnitude of the memory factor (q) of CADC increases fidelity and minimizes fidelity deviations. The weak measurements further help to improve the quality of QT. The increment in memory factor at the appropriate value of $w$ and $wr$ parameters, characterizing WM and RWM, gives better results. This work connects the impact of memory in quantum teleportation (QT) with the role of WM and RWM further improving its quality. Therefore, the memory effects are crucial and beneficial in improving QT. It is observed that high memory effects allow fidelity deviation values to reach zero even for non-zero time with WM and RWM.
We believe our analysis is beneficial for experimental QT and will be a good step toward realizing UQT. It also connects the impact of memory effects in QT, which can be verified experimentally. The role of WM and RWM in improving quantum entanglement is beneficial, as the quantum state with a concurrence value $\approx 0.19$ can be enhanced to a value close to 1. The shared entangled resource state can also be optimized using WM and RWM for UQT.

\begin{acknowledgments}
VBS, AK, and SB acknowledge support from the Indian Institute of Technology, Jodhpur to provide the facilities necessary to complete the work. S.B. acknowledges support from the Interdisciplinary Cyber-Physical Systems (ICPS) programme of the Department of Science and Technology (DST), India, Grant No.: DST/ICPS/QuST/Theme-1/2019/6. VBS acknowledges the Department of Chemistry, IIT Jodhpur and MoE for providing research facilities and financial support.
\end{acknowledgments}

\appendix*

\section{\label{app:all} Equations for states after WM and RWM protocol for unital and non-unital channels}
The Bell-type quantum state is undergoing the protocol presented in Fig. \ref{fig:wm_protocol}; we present the analytical form of the resulting final quantum state in the upcoming subsection. The evolution through noisy channels of unital and non-unital types gives the different forms of the final state. The explicit form of it is presented, respectively, for unital as well as non-unital channels. 

\subsection{\label{app:unital}For the case of unital channel acting on one of the qubit}
For the density matrix $\rho$ corresponding to a Bell-type state of the form $\ket{\psi}=\cos \theta \ket{00} + \sin \theta \ket{11}$,
the final state $\rho_{BT}^f(w,wr)$ after WM and RWM protocol, for the case of unital quantum channel affecting the travel qubit, is given by the following equation 
\begin{widetext}
\begin{equation}\label{final_state_unital}
   \begin{pmatrix}
\frac{(2 (1 - wr) \cos^2(\theta)}{(2 - wr - w - (wr - w) \cos(2 \theta)} & 0 & 0 & \frac{2 \sqrt{(1 - wr)} p(t) \sqrt{(1 - w)} \cos(\theta) \sin(\theta)}{(2 - wr - w - (pr - w) \cos(2 \theta)} \\
0 & 0 & 0 & 0\\
0 & 0 & 0 & 0\\
\frac{2 \sqrt{(1 - wr)} p(t) \sqrt{(1 - w)} \cos(\theta) \sin(\theta)}{(2 - wr - w - (wr - w) \cos(2 \theta)}  & 0 & 0 & \frac{(2 (1 - w) \sin^2(\theta)}{(2 - wr - w - (pr - w) \cos(2 \theta)}, 
\end{pmatrix}
\end{equation}
\end{widetext}
here $w$ are $wr$ are WM and RWM  strength of WM and RWM operations, respectively, and $p(t)$ is the noise function associated with unital channel under study.
\subsection{\label{app:non-unital}For the case of non-unital AD channel acting on one of the qubit}
For the density matrix $\rho$ corresponding to a Bell-type state of the form $\ket{\psi}=\cos \theta \ket{00} + \sin \theta \ket{11}$,
the final state $\rho_{BT}^f(w,wr)$ after WM and RWM protocol, for the case of non-unital AD channel affecting the travel qubit, is given by the following equation 
\begin{widetext}
\begin{equation}\label{final_state_unital}
   \begin{pmatrix}
\frac{((1 - wr) \cos^2(\theta)}{((1 - wr) \cos^2(\theta) + (1 - w) (1 - wr (p(t)) \sin^2(\theta)} & 0 & 0 & \frac{\sqrt{(1 - wr)} \sqrt{(1 - w)}\sqrt{(1 - p(t))}
  \cos(\theta) \sin(\theta)}{((1 - wr) \cos^2(\theta) + (1 - w) (1 - wr (p(t)) \sin^2(\theta)}\\
0 & 0 & 0 & 0\\
0 & 0 & \frac{(1 - wr)(1 - w) p(t)\sin^2(\theta)}{((1 - wr) \cos^2(\theta) + (1 - w) (1 - wr (p(t)) \sin^2(\theta)} & 0\\
\frac{\sqrt{(1 - wr)} \sqrt{(1 - w)}\sqrt{(1 - p(t))}
  \cos(\theta) \sin(\theta)}{((1 - wr) \cos^2(\theta) + (1 - w) (1 - wr (p(t)) \sin^2(\theta)} & 0 & 0 & \frac{(1 - wr) (1-p(t))\sin^2(\theta)}{((1 - wr) \cos^2(\theta) + (1 - w) (1 - wr (p(t))\sin^2(\theta)},
\end{pmatrix}
  \\
\end{equation}
\end{widetext}
here $w$ are $wr$ are the strength of WM and RWM operations, respectively, and $p(t)$ is the noise function associated with the non-unital AD channel under study.

\bibliography{final}

\end{document}